\documentclass[twocolumn,showpacs,preprintnumbers,amsmath,amssymb]{revtex4}

\usepackage{graphicx}
\usepackage{dcolumn}
\usepackage{bm}
\usepackage{epstopdf}
\usepackage[usenames,dvipsnames]{color}

\begin{document}


\title{Prices of Options as Opinion Dynamics of the Market Players with Limited Social Influence}
\author{Elad Oster}
\affiliation{The Hebrew University, Jerusalem, Israel}
\author{Alexander Feigel}
\email{alexfeigel@gmail.com}
\affiliation{Soreq NRC, Yavne 81800, Israel}


\date{\today}

\begin{abstract}

The dynamics of market prices is described as the evolution of opinions in the trading community regarding future market behavior. The price then is a function of the voting process of the market players in favor to raise or reduce the value of a stock. The model presented in this paper is suited for pricing of options and was verified against real market data. The model allows deriving the parameters of market players from available real market data, especially maximum possible correlation (herding) and anti-correlation between the players' opinions. The deviations of market prices from those predicted by the Black-Scholes model, such as smile and skew implied volatilities, are interpreted as the current values and limits of social influence of the market players, respectively. To the best of our knowledge, this is the first work that discriminates skew and smile phenomena. Our approach unifies and develops a further connection between trading, voters’ model, and statistical physics analogies of opinion dynamics.

\end{abstract}

\maketitle



\date{\today}

\maketitle

\section{Introduction}

The world of finance provides a challenge to the modeling of stock price dynamics as a function of the most important market characteristics. The prices of financial markets depend on the state of the economy, recent news, stock trading regulations, and especially the mood of the market players. To describe stock price dynamics, one should develop a united analytic framework for behavior of market players and for market mechanisms that translate this behavior into actual prices of a stock and its options (i.e., future contracts with a profit that depends on the specific behavior of the stock price in the future). Consequently, the modeling of stock markets is part of a more general problem of community opinion dynamics \cite{Castellano2009}\cite{Feigenbaum2003}.

The current prices of options are affected by demand and supply of the market players according to their beliefs regarding the future evolution of the prices. The future distribution of prices imagined by the market players might be revealed by the prices of the options\cite{BLACK1973}\cite{MERTON1973}. For instance, the greater expected change of the price of a stock, the greater the change in the price of the corresponding options that benefit from the raising market. Financial markets supply a huge amount of historical prices of stocks and their options to analyze, including phenomena such as nationwide financial crises.

An analysis of historical market prices clearly demonstrates abrupt changes that result in greater probabilities than in Gaussian (normal) distributions for large price steps. The Gaussian distribution is used as a reference because it is a cumulative result of many random independent contributions and it is widespread. Deviations from the normal distribution are known as fat-tail distribution\cite{Mandelbrot1963}. Moreover, the distribution of future prices imagined by the market players according to the options’ prices possesses the same fat-tail distribution. Non-Gaussian behavior of option prices is called volatility smile or skew\cite{Derman1994}. It is notable that the volatility smile developed significantly after the financial crisis of 1987\cite{Rubinstein1994}.

The abrupt changes of the prices indicate complex phenomena among the market players such as communication\cite{Eguiluz2000}, herding\cite{BANERJEE1992}\cite{Cont2000}, or crashophobia\cite{Rubinstein1994}. These phenomena result in a collective response of the market players that might explain abrupt (non-Gaussian) changes of the prices\cite{Stanley2008}\cite{Schinckus2013}\cite{Chakraborti2011}. Their connection to markets makes these phenomena of great practical importance.

Stock players are historically separated into bulls and bears that act in favor of growing or reducing the market, respectively. This analogy unites the problem of the market players with the models of opinion dynamics\cite{Sznajd-Weron2002}\cite{Cont2000}. Collective behavior such as herding emerges due to the social influence between the players\cite{BANERJEE1992}\cite{Vespignani2012}\cite{Kocsis2011}\cite{Eguiluz2000}\cite{Feigel2008}.

To test different models of voting or opinion dynamics against real data of the market prices, one should link the price of a stock with the amount of bulls and bears in the market players’ community. Previous attempts assumed a mere linear dependence of the price or change of the price on disbalance between the bulls and the bears in the community\cite{Sznajd-Weron2002}\cite{Cont2000}\cite{Farmer2002}\cite{Feng2012}. These assumptions generate prices with a fat-tail distribution in the case of herding of the market players. A comparison with real data and an explanation of the real phenomenon of volatility smile, however, require further clarification of the correct dependence between the stock price and the parameters of the market players, such as herding and social influence. 

This Article presents a general description of stock and options’ price formation as an opinion dynamics of the market players to be either bulls or bears. The model shows a good fit to real data of option prices, both of stocks and indices. According to our results, the non-Gaussian distribution of prices depends on social influence between the market players and the initial collective memory on distribution of bears and bulls. Moreover, social influence and collective memory discriminate two different patterns, skew and smile, of the option prices. The model describes both Gaussian and non-Gaussian distributions, allowing a discussion of how the latter might appear as it happened during the 1987 financial crisis. We argue that a vote to price connection may be independent of exact market mechanisms.

We demonstrate market modeling using the seminal Black-Scholes model for pricing options\cite{BLACK1973}\cite{MERTON1973}. This model relies on two major assumptions. The first is that the relative change in the price of a stock $dS/S$ in a time step $dt$ is the sum of the deterministic change in monetary value and a Gaussian stochastic process that describes contribution of the market players:
\begin{eqnarray}
  \label{eq:BSmain}
  \frac{dS}{S}=\mu dt+\sigma dz,
\end{eqnarray}
where $\mu$ is a prime rate, $dz=\phi(0,1)\sqrt(dt)$ is a Wigner process ($\phi(0,1)$ is a random variable of normal distribution with mean $0$ and standard deviation $1$). The parameter $\sigma$ is called volatility. 

The volatility $\sigma$ of the Black-Scholes model solely describes the result of the decision making and trades of the market players. The stochastic process (\ref{eq:BSmain}) corresponds to diffusion in $\log{S}$ space $d\log{S}=(\mu-\frac{1}{2}\sigma^2)dt+\sigma dz$ and, therefore, leads to a log normal distribution of the prices:
\begin{eqnarray}
  \label{eq:ln1}
  P_{BS}=\frac{1}{\sqrt{2\pi\sigma^2}}\exp{\frac{(\log{S}-\mu+\frac{1}{2}\sigma^2 t)^2}{2\sigma^2 t}}.
\end{eqnarray}
High values of volatility $\sigma$, therefore, correspond to greater changes of the stock prices at a given period of time. 

The second assumption of the Black-Scholes model is the non-arbitrage principle, also called an assumption of the effective market, which allows one to link the dynamics of the stock prices with the price of future contracts, for instance put or call options. Consider a plain vanilla call option, which is the right to buy a stock $S$ at some strike price $K$ after some maturity time $M$. The value of a call option $C$ with strike price $K$ as a function of the corresponding stock price $S$ at the maturity day is:
\begin{equation}
  C(S,K) = \max\left(0,S-K\right). \label{option0}
\end{equation}
This expression describes the possibility to buy stock at price $K$ and immediately sell it at price $S$. The value of the option is its average return multiplied by the expected change in monetary value until the maturity time:
\begin{equation}
  C_{BS}(\mu,\sigma,K,t_{M}) = \int_{0}^{\infty}P_{BS}(S,t_{M})C(S,K)dS, \label{option1}
\end{equation}
where $P(S,t)$ is the expected distribution of stock value at maturity time $t_{M}$. The distribution of stock value is unequivocally linked to the dynamics of the stock price. Any deviation from (\ref{option1}) leads to arbitrage possibilities between investments in stock and its options.

The Black-Scholes model describes the future distribution of stock prices $P_{BS}$ together with corresponding prices of the options $C_{BS}$ as the functions of the single free parameter, volatility $\sigma$, see eq. (\ref{option1}). One might, therefore, calibrate volatility against current prices of options using eq. (\ref{option1}). This calibration is essential to use $P$ for pricing exotic or non-tradable options and to get unobservable parameters such as $\frac{\partial C}{\partial S}$ for hedging. However, the calibration of the Black-Scholes model demonstrates that single parameter volatility $\sigma$ does not suffice to describe the dynamics of the observed market prices.

The real prices of options are presented by the implied volatility surface $\sigma_{imp}(K,t_{M})$, which describes the deviation of the reality from the Black-Scholes model\cite{Gatheral2006}. Implied volatility is defined for each strike price $K$ and maturity time $t_{M}$ as the volatility of the Black-Scholes model that corresponds to the current price of option $C_{obs}(K,t_{M})$:
\begin{eqnarray}
  \label{eq:rco}
  C_{BS}(\mu,\sigma_{imp}(K,t_{M}),K,t_{M})=C_{obs}(K,t_{M}),
\end{eqnarray}
where the Black-Scholes value of options is defined by eq. (\ref{option1}). The volatility surface is flat in case of a Black-Scholes market since eq. (\ref{option1} has the same solution $\sigma_{imp}(K,t_{M})=\sigma$ for all values of $K$ and $t_{M}$. In reality, the volatility surface is far from being flat, indicating a non-Gaussian behavior of the stock price.

After the financial crisis of 1987, implied volatility as a function of strike price $K$ for a given maturity time $t_{M}$ generally demonstrates smile or skew behavior (see Fig. \ref{fig1}). This deviation from flat Black-Scholes volatility indicates that stock price dynamics is not Gaussian. Skew is more present in indices that are averages of many stocks, and smile is more common for stocks.

To describe the volatility smile phenomenon, one should correct the stochastic process (\ref{eq:BSmain}) to include non-Gaussian dynamics. The main corrections include stochastic\cite{Hull1987}, local volatility\cite{Dupire1994}\cite{Derman1994}, and additional stochastic processes such as jumps or non-Gaussian behavior due to correlated behavior of the players (opinion aggregation)\cite{Gatheral2006}. The stochastic volatility approach assumes volatility $\sigma$ to be a stochastic process. Local volatility is the assumption that $\sigma(S,t_{M})$ is a function of stock price and maturity time. These two corrections possess high practical value. However, they lack reasonable justification and cannot explain the nature of the volatility smile phenomenon. Human parameters such as crashofobia or herding are difficult to quantify and to compare with real data. That leads to arguments regarding the true nature of the volatility smile and the corresponding dynamic of market players.

Market micromodeling using agent simulations is an alternative approach for deriving the stochastic process of a stock\cite{Bak1997}\cite{Samanidou2007}\cite{Wellman2004}\cite{Farmer2002}\cite{Muchnik2003}. A market player is modeled to the level of its strategy to buy and sell available assets. Market mechanisms then translate the actions of many players into changes of the prices. An advantage of this approach is a clear understanding of all processes on all levels. A disadvantage is the great number of parameters and the difficulty to define completely human trading as the actions of an agent in simulation.

This work presents price dynamics as a function of the limited number of parameters with special emphasis on social influence (herding) between the market players\cite{Sznajd-Weron2002}\cite{Cont2000}. We present an alternative way to network topology to define social responsivity that is better integrated in classic modeling of market prices by stochastic processes\cite{BLACK1973}. The advantages of our method include both positive and negative social influence, boundaries, and dynamics. We achieve a good fit of the model with real market data and answer some-long standing questions, such as the origin and especially the difference between smile and skew in implied volatility surfaces.  

When social influence is zero, our model converges to the Black-Scholes model result. This is due to the assumption that players without mutual effect compose a Gaussian Black-Scholes market.

The model is suitable for parallel Monte Carlo simulations. A Graphical Processing Unit (GPU) was used to calibrate the model's parameters to fit real data. The model may be optimized further for practical needs.

Next, we describe the model in detail, present the results of fitting real volatility surfaces, and discuss the basic assumptions, implications, and relations of our model to other works.

\begin{widetext}
\onecolumngrid
\begin{figure}
  \begin{center}
    \begin{tabular}{c c}
      \multicolumn{1}{l}{{\bf\sf A}} & \multicolumn{1}{l}{{\bf\sf B}} \\
      \resizebox{0.5\textwidth}{!}{\includegraphics{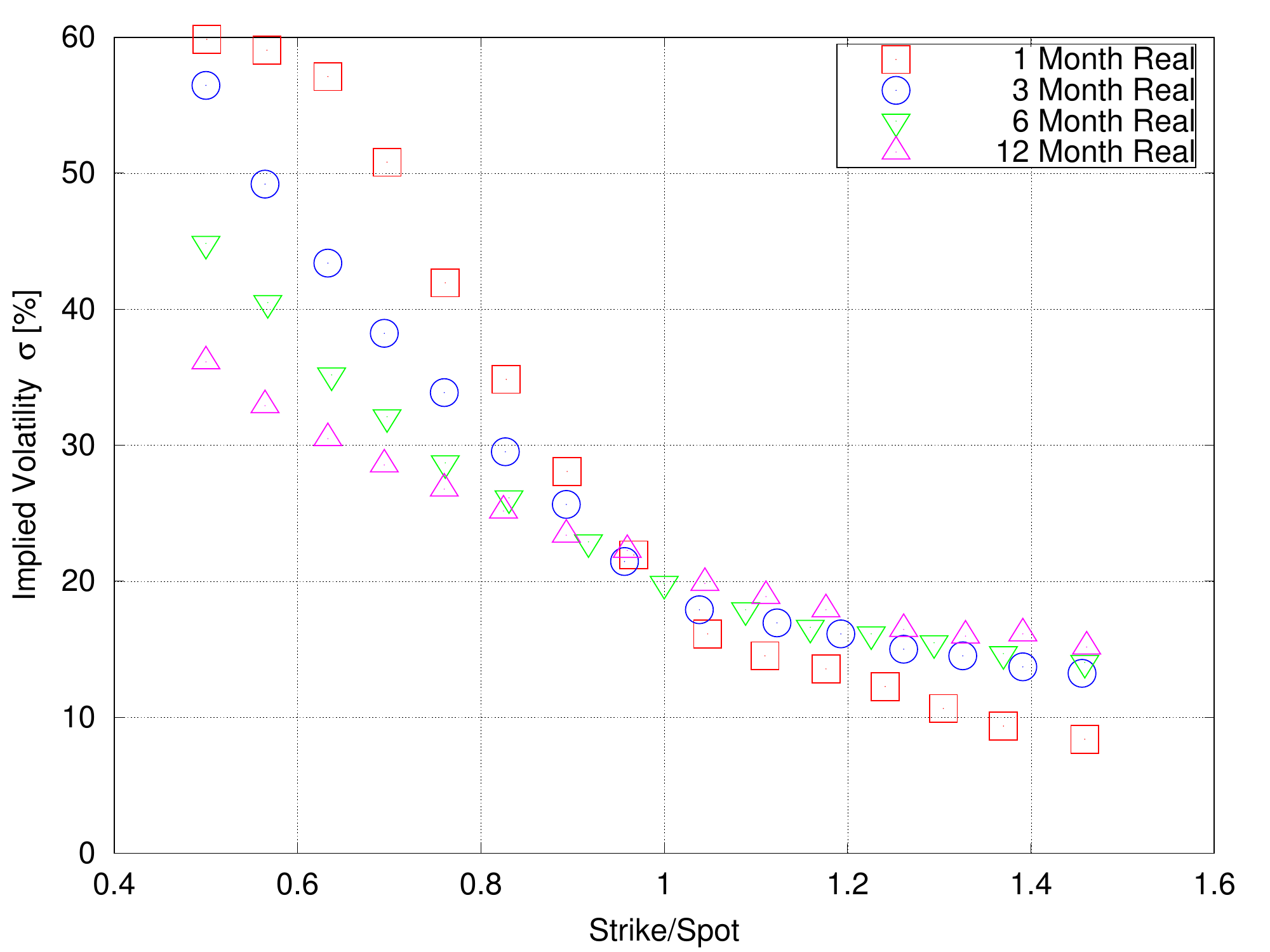}} &
      \resizebox{0.5\textwidth}{!}{\includegraphics{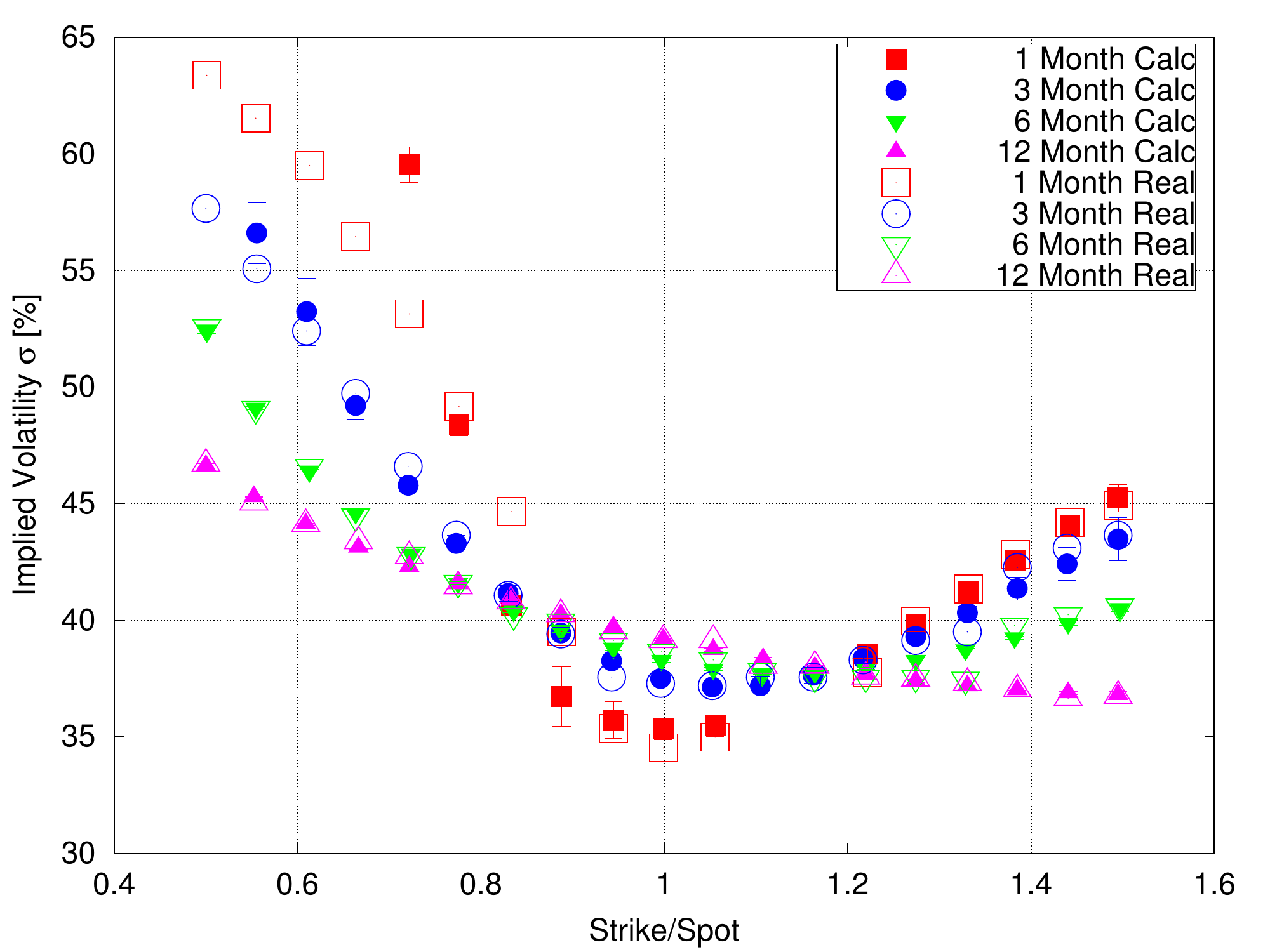}} \\ 
    \end{tabular}
    \caption{{\bf Implied volatility surfaces of skew and smile types.} Implied volatility of a given option is a function of the options’ strike price and maturity time. The implied volatility is presented at four maturity times (1,3,6, and 12 months) as a function of strike to spot prices ratio, where spot is the price of the corresponding stock or index at the time of the graph.  ({\bf A}) Implied volatility surface of SPX index at 12/26/01 with clear skew pattern. ({\bf B})  Vodafone Company (VOD) stock from and 12/27/01. It presents a more smile behavior than skew,  except late maturity times. The model of this work (solid markers) successfully fits the implied volatilities of both smile and skew types.}
    \label{fig1}
  \end{center}
\end{figure}
\twocolumngrid
\end{widetext}

\section{The Model}

\begin{figure}
\begin{center}
\resizebox{0.5\textwidth}{!}{\includegraphics{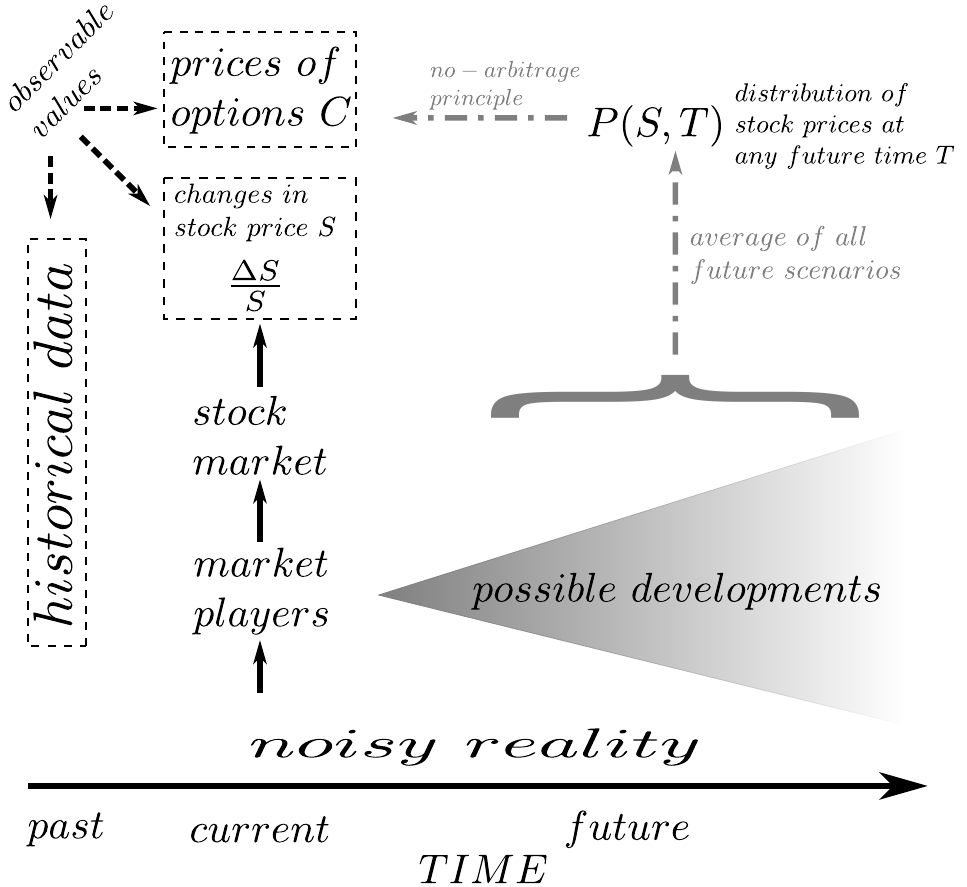}}
\caption{{\bf Stock market observable and hidden parameters.} The observable parameters of a market include historical and current prices of stocks, indices, and their options. The price of a stock $S$ is affected by noisy reality, its interpretation by the market players, and the market trading mechanisms that transform players’ actions into the price change $\Delta S$. The stochastic nature of the reality and market players results in a probability distribution $P(S,T)$ of the future stock price to possess a specific value $S$ in a succeeding time $T$. This distribution follows unequivocally from the prices of stock options $C$ (e.g., price of a permit to buy the stock in the future at a predetermined strike price) assuming an arbitrage-free efficient market. This distribution $P(S,T)$ is unobservable directly, though it is connected unequivocally with the prices of options through the non-arbitrage principle. A reliable market model has to predict the prices’ distribution $P(S,T)$ that corresponds to the observable values of current and historical option prices.}
\label{fig2}
\end{center}
\end{figure}

We model a market assuming the following general characteristics (see Fig \ref{fig2}). The market players collect available information and define their strategies to profit either from increasing or decreasing prices; the players that expect neutral market are neglected at this point. The players, according their choice, are called either bulls or bears. To decide their strategy, the players might observe historic and current prices of stocks together with the corresponding options. The market mechanisms translate the players’ decisions into real prices. The option prices depend on the non-observable distribution of future prices of the corresponding stock through non-arbitrage principle~(\ref{option1}).

There are two steps to defining the stochastic dynamics of the stocks as a function of the parameters of the market players. First, we describe trading as a voting process with two choices to be either bear or bull. Second, we derive the price change of a stock as a function of the outcome of the voting process. We argue that under general circumstances this function is unique, disregarding the exact market mechanisms. 

We can then derive option prices as a function of the parameters of the market players’ community. The distribution of the future stock prices is obtained by averaging stochastic stock price dynamics. It allows calculating the corresponding options prices and calibrating the model parameters against real market data.

\subsection{Market of non-interacting players and Black-Scholes (Gaussian) prices}

\begin{figure}
\begin{center}
\resizebox{0.5\textwidth}{!}{\includegraphics{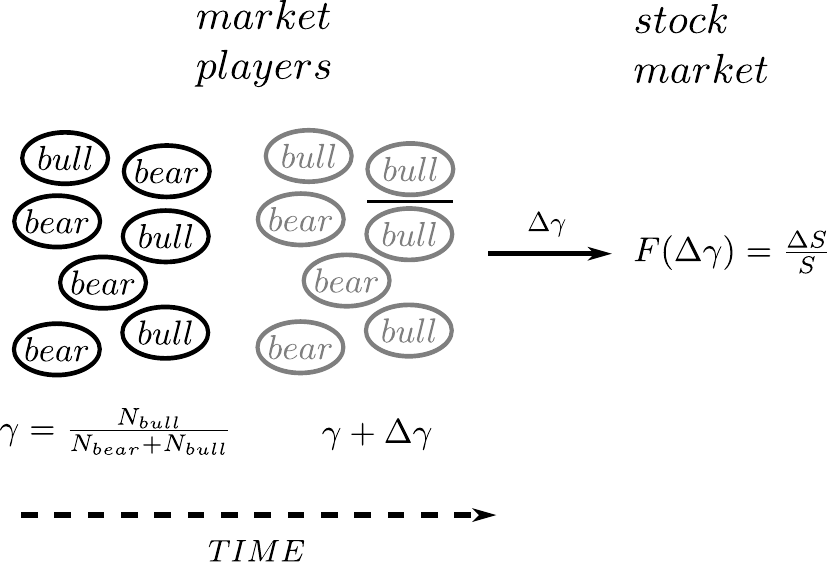}}
\caption{{\bf Stock price dynamics as a voting process by the market players.} The market players are generally separated into bulls and bears. Bulls play long: they buy and sell stocks, indices, and their option to make profit from the raising market. Bears, on the other hand, play short: they try to profit from decreasing prices. The number of bulls and bears among the market players changes with time. It resembles a voting process where at each time a player has to decide either to be a bull (vote up in favor of a raising market) or to be a bear (vote down against a raising market). In this article, we hypothesize that this process defines the dynamics of stock prices. The market then is presented as a function $F(\Delta\gamma)$, where $\Delta\gamma$ is the change in ratio between bulls and total number of the market players.}
\label{fig3}
\end{center}
\end{figure}

According to this model, the market players are separated into bulls that push the prices up and bears that try to move the market down. The ratio between bulls and all market players $\gamma$ is:
\begin{eqnarray}
  \label{eq:gamgam}
  \gamma = \frac{N_{bull}}{N_{bull}+N_{bear}}=\frac{N_{bull}}{N},
\end{eqnarray}
where $N_{bull}$ and $N_{bear}$ are the total numbers of bulls and bears, respectively, and $N$ is the total number of players. The ratio $\gamma$ changes with time since the players change their state occasionally (see Fig. \ref{fig3}). The trade advances by discrete steps in time.
 
To derive the stochastic process for the ratio of bulls to the total number of market players $\gamma$ during a single trading step, we make two assumptions. First, the number of transactions is proportional to the number of possible interactions between bulls and bears $\gamma(1-\gamma)$. Thus, $\gamma$ changes either by a positive or by a negative step:
\begin{eqnarray}
  \label{eq:dfgr2}
  \Delta\gamma_\pm\propto\pm\gamma(1-\gamma). 
\end{eqnarray}
Second, in the absence of communication, the probability of a player to change its state (from bear to bull or vice versa) is assumed equal for all players. Consequently, the probabilities of positive and negative steps are:
\begin{eqnarray}
  \label{eq:dfgfr4}
  &&P_+=1-\gamma,\\\nonumber
  &&P_-=\gamma, 
\end{eqnarray},
respectively, because $\gamma$ is the ratio of bulls in the community (\ref{eq:gamgam}). 

Following (\ref{eq:dfgr2}) and (\ref{eq:dfgfr4}), the stochastic dynamics of bull ratio as a generalized Wigner process:
\begin{equation}
 d\gamma = \mu\left(\gamma,t\right) dt+\sigma\left(\gamma,t\right) dz_\gamma, \label{Gauss1}
\end{equation}
taking into account that $\sigma\propto \sqrt{(\Delta\gamma_+)^2P_++ (\Delta\gamma_-)^2P_-}$ and $\mu\propto\Delta\gamma_+P_++ \Delta\gamma_-P_- $, is: 
\begin{equation}
  d\gamma = -(\gamma-\gamma^\star)\gamma(1-\gamma)\sigma_{\gamma}^{2}dt+\sigma_{\gamma}\gamma(1-\gamma)dz_\gamma, \label{gam1}
\end{equation} 
where $\sigma_{\gamma}$ is an arbitrary constant and $\gamma^\star=1/2$. The main properties of (\ref{gam1}) are the limited range of $0<\gamma<1$ and zero drift $\mu(\gamma,t)=0$ term in the case of  neutral population $\gamma=1/2$. The latter is a direct consequence of the second assumption, i.e., an unbiased population (equal probability for all players). In general, biased population equilibrium occurs at different values $0<\gamma^\star<1$, because the step probabilities become $P^\star_+=0.5+\gamma^\star-\gamma$ and  $P^\star_-=0.5-\gamma^\star+\gamma$. 

Having a stochastic process for $\gamma$ (\ref{gam1}), we are now searching for a market function $F(\gamma)$ that translates the voting process into stock price dynamics of the Black-Scholes type: 
\begin{equation}
  dF(\gamma) = \frac{dS}{S} = \mu_{S}dt+\sigma_{S}dz_{S}. \label{stock0}
\end{equation}
To find this function, we use Ito’s lemma, describing the differential of a time-dependent function of a stochastic process (\ref{Gauss1}):
\begin{equation}
dF(\gamma,t) = \left (\frac{\partial F}{\partial\gamma}\mu+\frac{\partial F}{\partial t}+\frac{1}{2}\frac{\partial^{2}F}{\partial\gamma^{2}}\sigma^{2} \right )dt+\sigma\frac{\partial F}{\partial\gamma}dz. \label{Ito}
\end{equation}
The requirement (\ref{stock0}) means that coefficients of both terms in (\ref{Ito}) are constant, similar to Eq.~(\ref{stock0}). 

Following (\ref{stock0}) and (\ref{Ito}), the market function $F(\gamma)$ that translates the voting process (\ref{gam1}) into Black-Scholes stock price dynamics is:
\begin{equation}
  F(\gamma)=B\log\frac{\gamma}{1-\gamma}. \label{F0}
\end{equation}  
where $B$ is a numeric coefficient. Substituting (\ref{F0}) in (\ref{Ito}), we get:
\begin{equation}
  dF(\gamma)=B\sigma_{\gamma}^{2}\left(\gamma^{\star}-\frac{1}{2}\right)dt+B\sigma_{\gamma}dz_F, \label{F1}
\end{equation}
The change in the price of a stock as a function of $\gamma$ is:
\begin{equation}
\frac{dS}{{S}}=B\Delta\log\left(\frac{\gamma}{{1-\gamma}}\right), \label{stock1}
\end{equation}
Integration of (\ref{stock1}) gives:
\begin{equation}
S=\left (\frac{\gamma}{1-\gamma} \right )^B. \label{stock111}
\end{equation}
It follows from analogy between (\ref{F1}) and (\ref{stock0}) if:
\begin{eqnarray}
  \label{eq:sim1}
  &&\mu_S\propto B\sigma_\gamma^2\left (\gamma^\star-\frac{1}{2}\right ),\\\nonumber
  &&\sigma_S\propto B\sigma_\gamma.
\end{eqnarray}
and assuming that: 
\begin{equation}
  dz_{\gamma}=dz_{S}, \label{dz0}
\end{equation}  
the noise of a stock is the noise of the voting process and the market is deterministic in the sense that it does not contribute with additional noise. 

The stochastic process for the ratio of bulls to the total number of the market players (Eq.~\ref{gam1}) should be modified to include the deviation of stock price dynamics from the Gaussian process (eq.~\ref{stock0}). We argue that the expression Eq.~\ref{stock1} holds for any modification of Eq.~\ref{gam1}. It is true if the market is described by a function of the vote outcome $\gamma$ and is independent of how it was obtained.

\subsection{Market players with social influence}
 
To extend the model to include stock price dynamics that is different from the Black-Scholes model (\ref{stock0}), we introduce the social influence between the market players. Social influence means that the state of a player depends on the state of the other players rather than being bull or bear with probabilities $\gamma$ and $1-\gamma$ independent of the environment.

We describe the interaction between market player $i$ and any other randomly selected player $j$ as follows: The probability per contact of player $i$ to be bear ($P_{bear}^{ij}$) depends on the state of player $j$. This conditional probability is given by:
\begin{equation} \label{eq:conditional probability 1}
P_{bear}^{ij} =
\begin{cases} 
	\alpha_{ij} & \mbox{if } s_j = 1 \\
	\beta_{ij}  & \mbox{if } s_j = 0 
\end{cases} 
\quad
= \alpha_{ij} s_j + \beta_{ij} (1-s_j),
\end{equation}
where $s_j$ is the state of player $j$ ($s_j = 1$ for bull state and $s_j = 0$ for bear state) and parameter $\alpha_{ij}$ and $\beta_{ij}$ is the probability per contact of player $i$ being bear given player $j$ is bear or bull correspondingly, regardless of the state of player $i$ prior to the interaction with player $j$.

\begin{figure}
	\resizebox{0.5\textwidth}{!}{\includegraphics{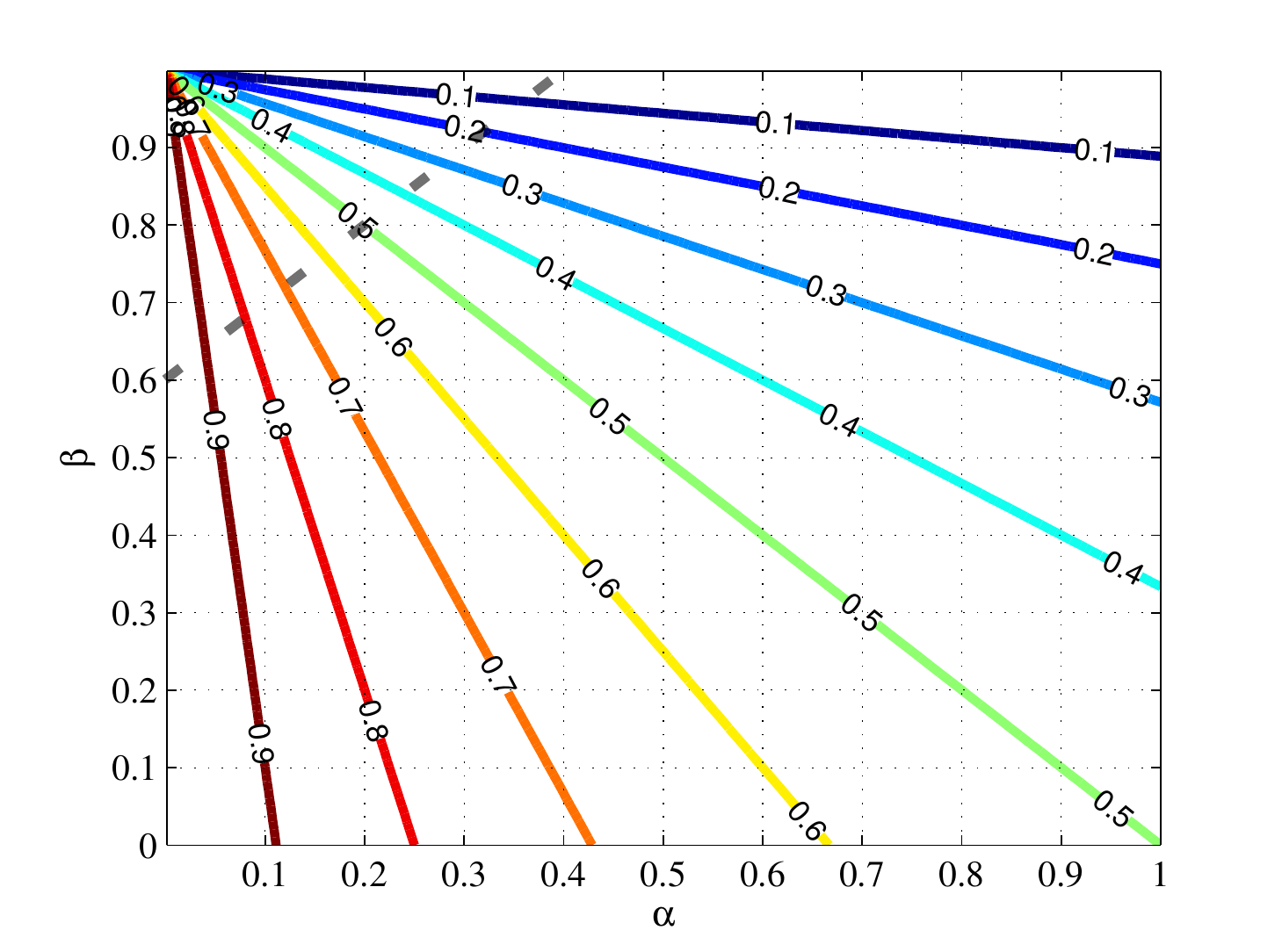}}
	\caption{ {\bf Population of market players: the average state.} The average amount of bulls $\gamma$ in a population composed only of the players $(\alpha,\beta)$ is obtained using mean field methods. The same values of $\gamma$ may correspond to different probabilities of player market correlations. These are defined by $(\alpha,\beta)$. For instance, $\gamma=0.5$ for both correlated $(\alpha=1,\beta=0)$ and anti-correlated $(\alpha=0,\beta=1)$ states, see eq. (\ref{omegaAB}). The sensitivity of $\gamma$ is greater near the correlated state  $(\alpha=0,\beta=1)$. The dashed line indicates the $I=0.6$ herding parameter.}
	\label{fig:gamma}
\end{figure}

In the mean field approximation, players exposed equally to the state of all other players ($\alpha_{ij},\beta_{ij})\equiv(\alpha,\beta$), Eq.~\ref{eq:conditional probability 1} become
\begin{equation} \label{eq:conditional probability 3}
P_{bear} \equiv P_{bear}^{i} = \frac{1}{N} \sum_{j=1}^N {\left[\alpha s_j + \beta (1-s_j)\right]}.
\end{equation}
Using the definition for $\gamma = P_{bull} = 1-P_{bear}$, we get
\begin{equation} \label{eq:Pdown}
P_{bear}=1-\gamma=\gamma\alpha + (1-\gamma)\beta.
\end{equation}
Therefore, we can derive an expression for $\gamma$ in terms of the conditional probabilities $\alpha$ and $\beta$
\begin{eqnarray}
\label{eq:mean-field-gamma}
\gamma=\frac{1-\beta}{1+\alpha-\beta},
\end{eqnarray}
because in steady state the ratio of bull players to the total number of the players equals to the probability of players to be bull (see Fig. \ref{fig:gamma}).

To calculate social influence\cite{Oster2015a}, let us estimate the response of the homogeneous community $(\alpha,\beta)$ to the injection of a group of relative size $\rho$ and unconditional average state $\gamma_\rho$. An unconditional response is independent of other players. The mean field eq. (\ref{eq:Pdown}) in this case becomes:
\begin{eqnarray}
  \label{eq:wf2}
  1-\gamma &=& (1-\rho)(\gamma\alpha+(1-\gamma)\beta)+\\\nonumber
  &&+\rho(\gamma_\rho\alpha+(1-\gamma_\rho)\beta),
\end{eqnarray}
because a player possesses probabilities $1-\rho$ and $\rho$ to interact with a responsive and an unconditional player, respectively. Eq. (\ref{eq:wf2}) can be rewritten as:
\begin{eqnarray}
  \label{eq:wf1}
  \gamma &=& \gamma(1-\alpha)+(1-\gamma)(1-\beta)+\rho\Delta\gamma(\beta-\alpha),
\end{eqnarray}
where $\Delta\gamma=\gamma_\rho-\gamma$ and, therefore, $\rho\Delta\gamma$ indicates the strength of the injection. 

The responsivity of average ratio of bulls to the total number of market players $\gamma$ to the injection of the unconditional group of strength $\rho\Delta\gamma$, following (\ref{eq:wf1}), is:
\begin{eqnarray}
  \label{eq:wf3}
  \chi_s=\frac{\partial \gamma}{\partial (\rho\Delta\gamma)}=\frac{\beta-\alpha}{1-(\beta-\alpha)},
\end{eqnarray}
For small perturbations $\rho\Delta\gamma$ one gets:
\begin{eqnarray}
  \label{eq:4f5g}
  \gamma_{final}=\gamma_{initial}+\chi_s\rho(\gamma_\rho-\gamma_{initial}), 
\end{eqnarray}
where $\gamma_{initial}$ and $\gamma_{final}$ are the average ratios of bulls prior and after the perturbation occurs. Social responsivity vanishes $\chi_s=0$ if $\alpha=\beta$ because in this case the players possess probabilities $\gamma$ to be bear independent of the state of the other players. Responsivity diverges at $(\alpha=0,\beta=1)$. At this point, a single individual’s change of state causes a phase-like transition of the state of the entire community.

Social responsivity $\chi_s$ (\ref{eq:wf3}) depends on the single herding parameter:
\begin{eqnarray}
  \label{eq:ii34}
  I=\beta-\alpha,
\end{eqnarray}
with range $-1<I<1$. It might be called herding because the contribution of $\rho\Delta\gamma$ unconditional group to $\gamma$ of responsive players is proportional to $I$; see (\ref{eq:wf1}). Thus, it indicates how much the responsive players are affected by a single member of the injected unconditional group. For comparison, the herding parameter is defined in \cite{Cont2000} using community graph topology as a ratio $c$ of the community that are connected to a single member and form a cluster of correlated behavior. The main advantage of $I$ is that it describes both positive and negative social influence. 

The absolute value of the herding parameter $|I|$ might be limited in real communities because high social responsivity assumes a high level of direct or indirect information exchange in community. A high level of information flow is limited because of the stochastic nature of market news, the topology of communication channels, and the tendency to hide individual strategies. 

The mutual information of the players can be a measure of information exchange in the trading community. Mutual information between market players $A$ and $B$ is the amount of known information about state of $B$ if state of $A$ is known for certain. Greater mutual information indicates either direct or indirect information flow. The direct information flow assumes communication, while the indirect information flow assumes synchronization by common external signal. Mutual information makes possible to separate markets with completely random and highly correlated choices of the players, corresponding to the lack of social influences and developed social influences, respectively.

\begin{figure}
\begin{center}
\resizebox{0.5\textwidth}{!}{\includegraphics{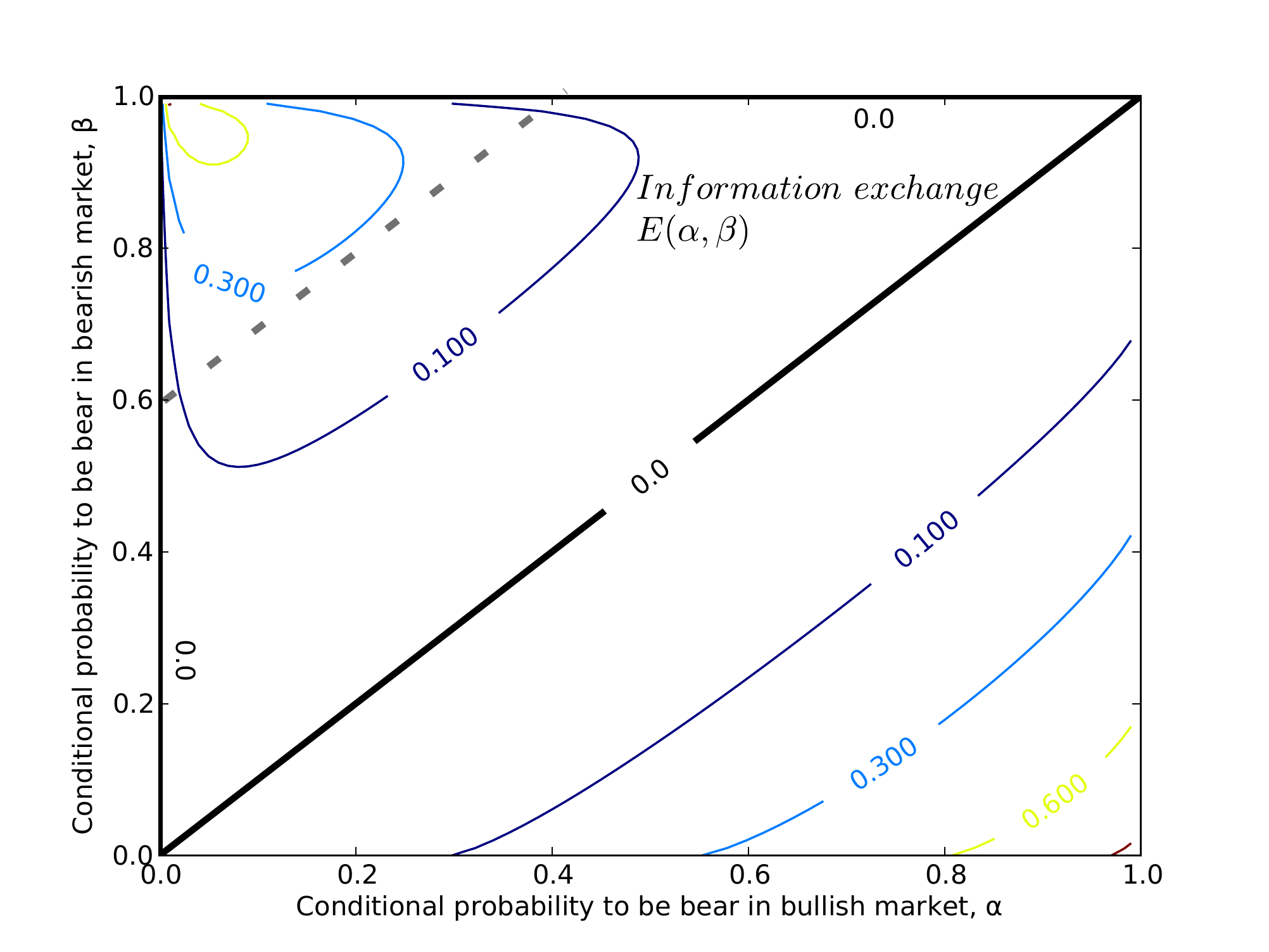}}
\caption{{\bf Mutual information and herding of identical market players $(\alpha,\beta)$.} The value of this mutual information describes the amount of information that is known about the state of the market provided by a state of the player and vice versa. Mutual information is maximum $=1bit$ for populations that correlated $(\alpha=0,\beta=1)$ or anti-correlated $(\alpha=0,\beta=1)$ with the markets. In the case of random $\alpha=beta$ or constant responses $\alpha=0$ or $\beta=1$, the mutual information is $=0bit$. The finite value of mutual information, therefore, is impossible without either direct or indirect exchange of information between the players. The dashed line indicates $I=0.6$ herding parameter.}
\label{fig5}
\end{center}
\end{figure}

Mutual information in a community of homogeneous players $(\alpha,\beta)$, see Fig. \ref{fig5}, is defined by probabilities $\Omega_{Bull,Bull}$, $\Omega_{Bull,Bear}$, $\Omega_{Bear,Bull}$ and $\Omega_{Bear,Bear}$ for four possible interactions in the community:
\begin{eqnarray}
E&=&\sum_{p,q=Bull,Bear}\Omega_{p,q}\times\nonumber \\
\times&\log_{2}&\left (\frac{\Omega_{p,q}}{{(\Omega_{p,Bull}+\Omega_{p,Bear})(\Omega_{Bull,q}+\Omega_{Bear,q})}}\right ).\nonumber \\\label{SMmutinfo1}
\end{eqnarray}
These probabilities are:
\begin{eqnarray}
\Omega_{Bull,Bull}=(1-\alpha)\gamma,\;\Omega_{Bear,Bear}=\beta(1-\gamma) \nonumber \\
\Omega_{Bear,Bull}=\alpha\gamma,\;\Omega_{Bull,Bear}=(1-\beta)(1-\gamma),
\label{omegaAB}%
\end{eqnarray}
following the definition of conditional probabilities $\alpha$ and $\beta$. Substituting (\ref{omegaAB}) in (\ref{SMmutinfo1}) we get:
\begin{eqnarray}
E(\alpha,\beta) &=&\nonumber \\
=&\log_{2}&\left (\frac{(1-\alpha)^{(1-\alpha)\gamma}(1-\beta)^{(1-\beta)(1-\gamma)}\alpha^{\alpha\gamma}\beta^{\beta(1-\gamma)}}{{\gamma^{\gamma}(1-\gamma)^{1-\gamma}}}\right ),\nonumber \\
\label{mutinfo2}
\end{eqnarray}
where $\gamma$ is defined by (\ref{eq:mean-field-gamma})\cite{Feigel2008}. 

The relationship between herding parameter $I$, ratio of bulls $\gamma$ (\ref{eq:mean-field-gamma}), and mutual information $E$ (\ref{mutinfo2}) is presented in Figs. \ref{fig:gamma} and \ref{fig5}.  Communities with the same herding $I$ might possesses arbitrary values of $\gamma$. The specific value of herding, however, constraints the possible maximum value of mutual information in the community and, vice versa, the specific value of mutual information limits the possible herding. For instance, the mutual information in community $\gamma=0.5$ is:  
\begin{eqnarray}
  \label{eq:IMref}
  E=\log_2\left [(1+I)^\frac{1+I}{2}(1-I)^\frac{1-I}{2} \right],
\end{eqnarray}
These constraints are valid for social influence (\ref{eq:wf3}), making impossible to achieve the maximum value at $I=1$ ($\alpha=0,\beta=1$ point). 

\subsection{General Market Model}
\label{sec:general-market-model} 

The community of the market players is considered as a point in the space of conditional probabilities $(\alpha,\beta)$, see Fig.~\ref{fig7}. The dynamics is limited to a subspace that is bounded by two mutual influence limits, namely $I_{low} < I < I_{up}$. On the mutual influence boundaries, we have reflecting boundary conditions.

If $I_{low},I_{up}\approx 0$ then $\alpha\approx\beta$, there is no mutual influence and the players’ state is independent of other players. In this case, the model converges to the Black-Scholes result, since the dynamics is described by asymmetric random walk (Eq.~\ref{gam1}) and the stocks’ price exhibit a log-normal behavior (definition of $F(\gamma)$). 

In the general case $I_{low},I_{up}\neq 1$, the community propagates by the stochastic process:
\begin{eqnarray}
  d\alpha&=& -(\alpha-\alpha^\star)\alpha(1-\alpha)\sigma_{\alpha}^{2}dt+\sigma_{\alpha}\alpha(1-\alpha)dz_\alpha, \\\nonumber
  d\beta&=& -(\beta-\beta^\star)\beta(1-\beta)\sigma_{\beta}^{2}dt+\sigma_{\beta}\beta(1-\beta)dz_\beta, \label{alpbet}
\end{eqnarray} 
starting from some initial point $(\alpha_0,\beta_0)$. It is derived assuming that, if the market is fixed to either bullish or bearish state, the ratio of bears ($\alpha$ and $\beta$, respectively) behaves analogous with $\gamma$ (Eq.~\ref{gam1}). Moreover, in the case of random responses $\alpha=\beta$, the process (Eq.~\ref{alpbet}) should exactly converge to (Eq.~\ref{gam1}) taking into account (Eq.~\ref{eq:mean-field-gamma}). 

The parameters $\alpha^\star$ and $\beta^\star$ in (\ref{alpbet}) indicate equilibrium values of $\alpha$ and $\beta$ in case noise term vanishes. The choice of $(\alpha^\star,\beta^{\star})$ is ambiguous. In this work, we assume that this point corresponds to neutral population $\gamma=0.5$ and possesses the same herding coefficient $I$ as the current state $(\alpha,\beta)$ of the trading community:

\begin{eqnarray}
\label{stars}
  \beta^{\star} &=& \frac{1}{2}+\frac{1}{2}(\beta-\alpha),\\\nonumber
  \alpha^{\star} &=& \frac{1}{2}-\frac{1}{2}(\beta-\alpha),
\end{eqnarray}
The sensitivity to specific values of $(\alpha^\star,\beta^\star)$ on our results is low; it was checked by assuming $\alpha^\star=\beta^\star=1/2$. 

The stock price $S$ is updated at each time step of the process. The initial value is set to $S=1$. The simulation proceeds by steps in time of arbitrary small value $\Delta t$. At each step, the next coordinates $(\alpha_{next},\beta_{next})$ are calculated:

\begin{eqnarray}
  \label{eq:abab}
\Delta\alpha_\mu &=& -(\alpha-\alpha^\star)\alpha(1-\alpha)*\Delta t*\sigma_\alpha^2,\\\nonumber
\Delta\alpha_\sigma &=& \alpha(1-\alpha)*\Delta t^{0.5}*\sigma_\alpha*\phi(0,1),\\\nonumber
\alpha_{next} &=& \alpha+\Delta\alpha_\mu+\Delta\alpha_\sigma.  
\end{eqnarray}

\begin{eqnarray}
  \label{eq:abab1}
\Delta\beta_\mu &=& -(\beta-\beta^\star)\beta(1-\beta)*\Delta t*\sigma_\beta^2,\\\nonumber
\Delta\beta_\sigma &=& \beta(1-\beta)*\Delta t^{0.5}*\sigma_\beta*\phi(0,1),\\\nonumber
\beta_{next} &=& \beta+\Delta\beta_\mu+\Delta\beta_\sigma.  
\end{eqnarray} 
where $\alpha^\star$ and $\beta^\star$ are defined by (\ref{stars}). New coordinates $(\alpha_{next},\beta_{next})$ lead to a new value of $\gamma_{next}$:
\begin{eqnarray}
  \label{eq:ggg}
  \gamma_{next} = \frac{1-\beta_{next}}{1-\beta_{next}+\alpha_{next}} ,
\end{eqnarray}
following Eq.~\ref{eq:mean-field-gamma}. The new stock price is calculated using Eq.~(\ref{stock1}):
\begin{eqnarray}
  \label{eq:sss}
S_{next} = S+S*\left [ \log\frac{\gamma_{next}}{1-\gamma_{next}}-\log\frac{\gamma}{1-\gamma}\right ]*B_{par},  
\end{eqnarray}
Then, call option prices are calculated for different strike prices $K$
\begin{eqnarray}
  \label{eq:cpr}
  C(t,K) = \max(S(t)-K), 
\end{eqnarray}
where $C(t,K)$ is an average over different runs for a set of times and strike prices.

Finally, implied volatility is calculated solving Eq.~\ref{eq:rco} by iterations. The prime value $\mu$ is used only while calculating implied volatility. The prime is neglected in the voting process itself. This volatility value is reduced by \emph{constant} $\Delta_{market}$
\begin{equation}
  \sigma_{imp}=\sigma_{players}-\Delta_{market}. \label{offset0}%
\end{equation}
This new parameter $\Delta_{market}$ essential to fit the real values of volatility. It can be interpreted as a reduction of volatility by market regulations.

\begin{widetext}
	\onecolumngrid
	\begin{figure}
		\begin{center}
			\begin{tabular}{c c}
				\multicolumn{1}{l}{{\bf\sf A}} & \multicolumn{1}{l}{{\bf\sf B}} \\
				\resizebox{0.5\textwidth}{!}{\includegraphics{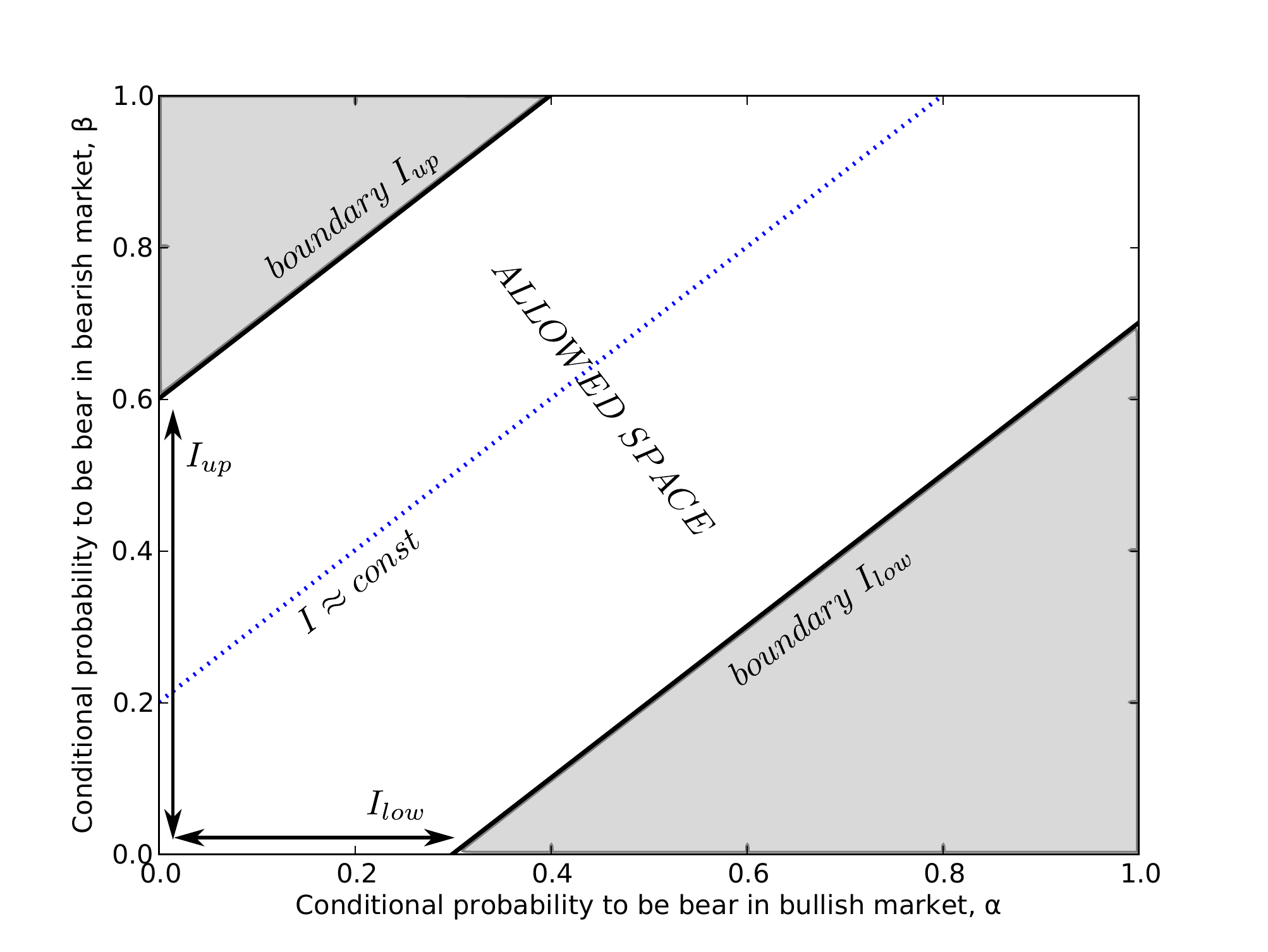}} &
				\resizebox{0.5\textwidth}{!}{\includegraphics{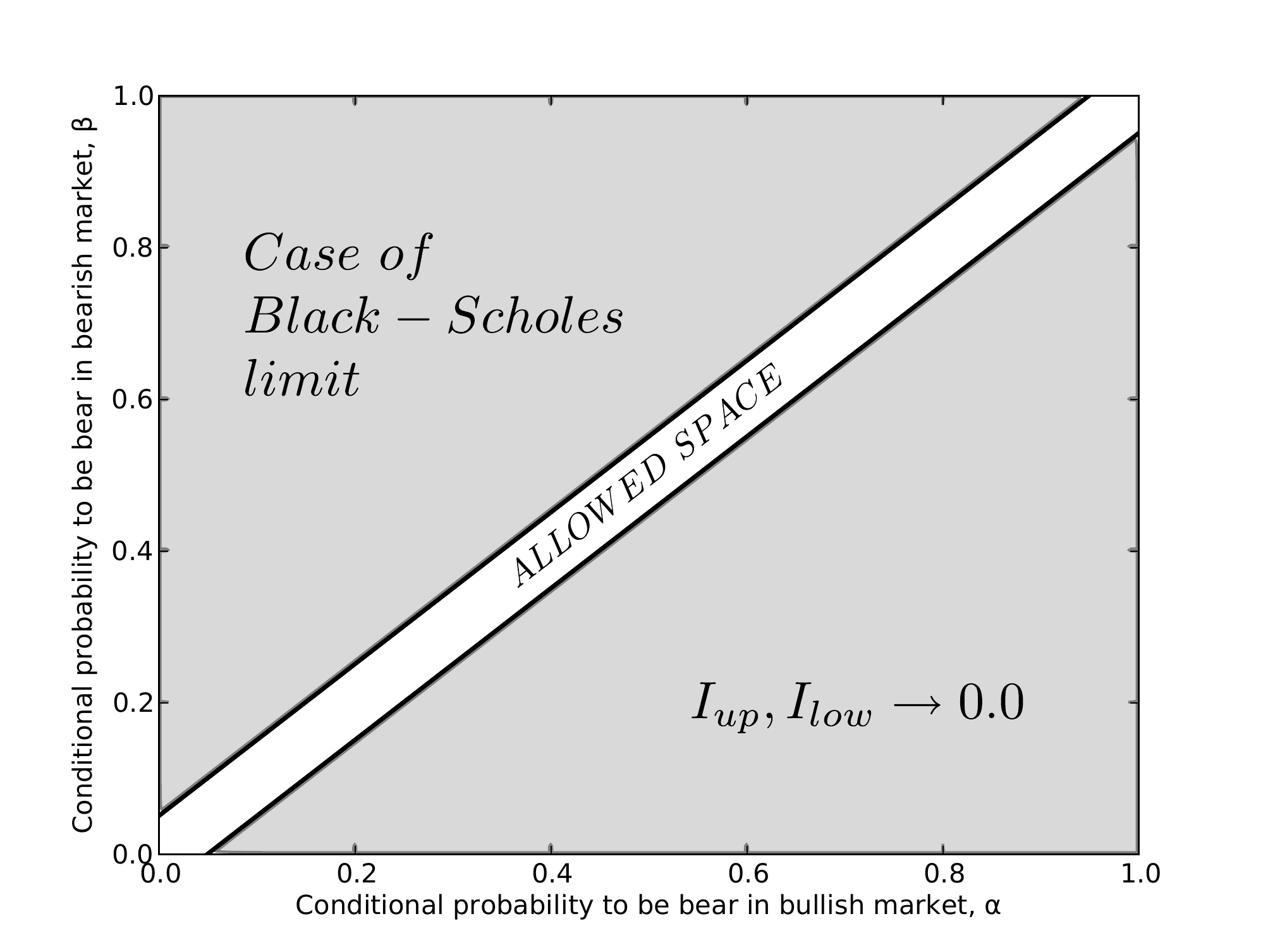}} \\
				\multicolumn{1}{l}{{\bf\sf C}} & \multicolumn{1}{l}{{\bf\sf D}} \\
				\resizebox{0.5\textwidth}{!}{\includegraphics{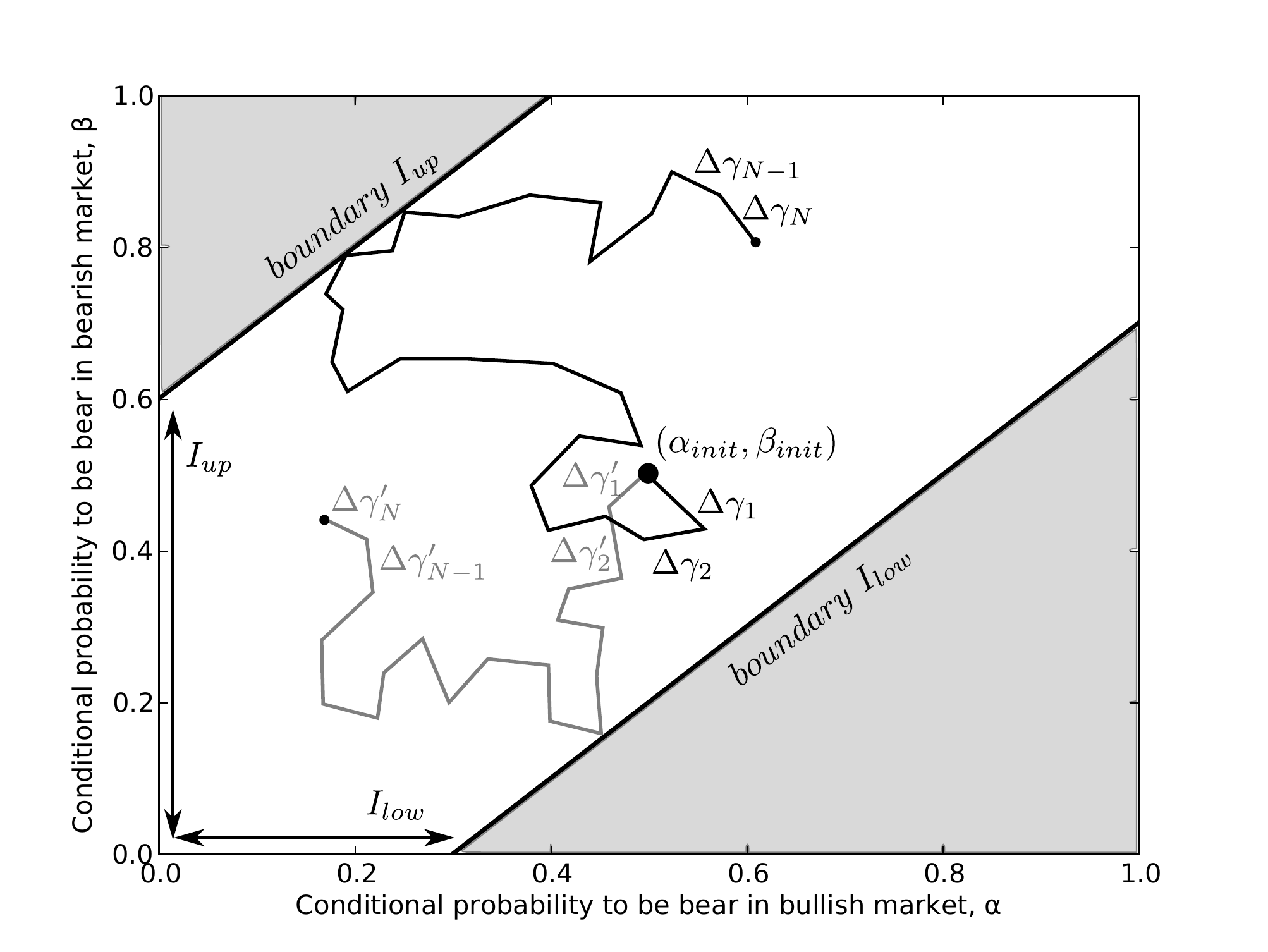}} &
				\resizebox{0.5\textwidth}{!}{\includegraphics{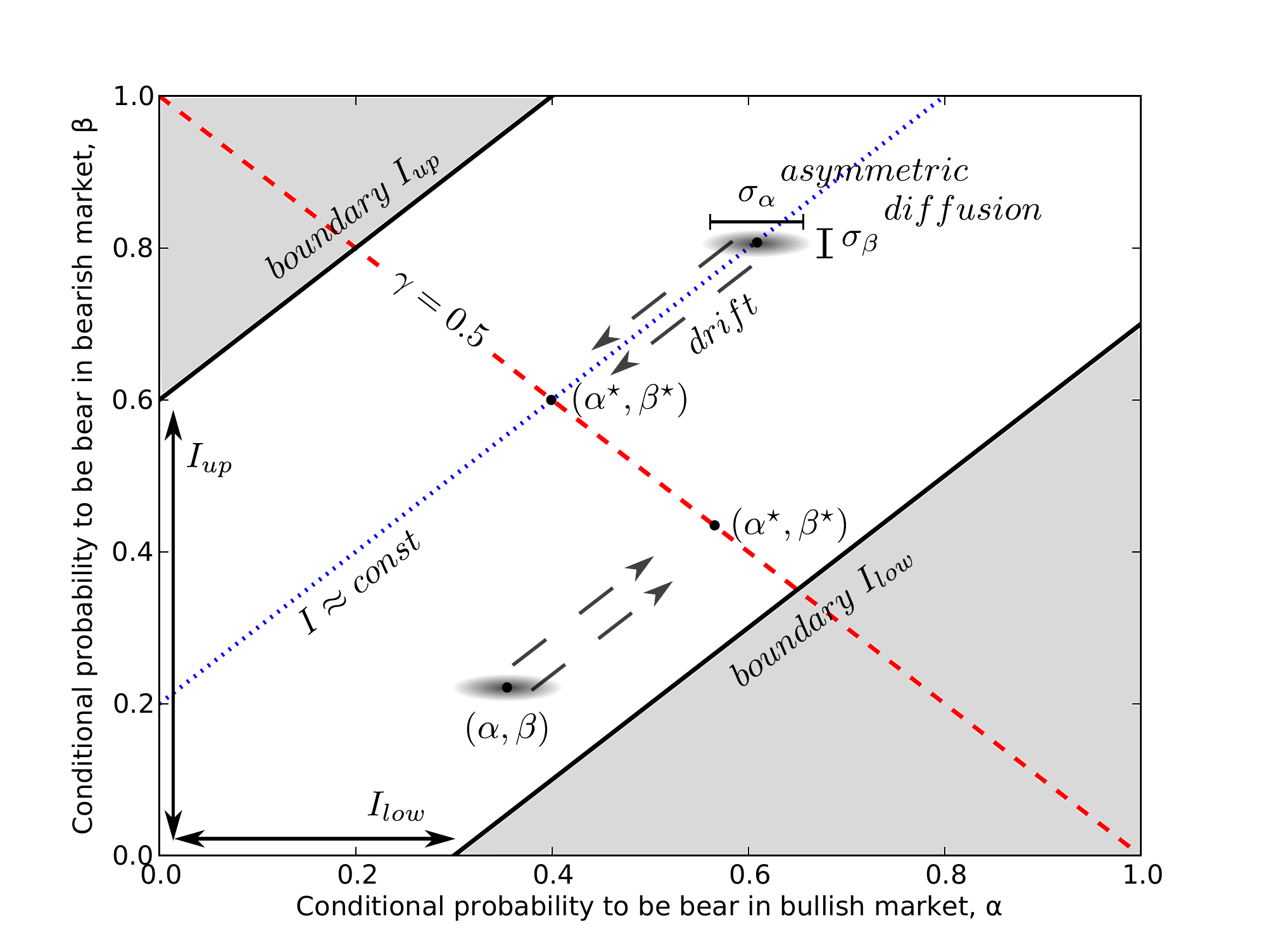}} \\
			\end{tabular}
			\caption{{\bf The dynamics of the market players in conditional probabilities space and corresponding dynamics of stock price} The dynamics of market players’ population in $(\alpha,\beta)$ space defines the dynamics of the corresponding stock price $S$. The population is described by a single point $(\alpha,\beta)$. The dynamics of the population is the diffusion with a drift under topological constraints. ({\bf A}) Topological constraints are defined by imposing limits on herding $I$ in the community $\beta = \alpha+I_{up,low}$. ({\bf B}) The limit of the Black-Scholes model occurs in the case of $I_{up},I_{low}\approx 0$, which limits the population to random behavior $\alpha\approx\beta$. ({\bf C}) The population starts at the initial position $(\alpha_0,\beta_0)$. Each step of the population is transferred to $\Delta\gamma$ and then to $\Delta S$ using eq. (\ref{stock1}). ({\bf D})  The population is subjected to diffusion, which is asymmetric along $\alpha$ and $\beta$ axis. We assume the drift if $\sigma_{\alpha}=\sigma_{beta}=0$ population converges to $\gamma=0.5$ while preserving its value of herding $I$.}
			\label{fig7}
		\end{center}
	\end{figure}
	\twocolumngrid
\end{widetext}

The model’s parameters are summarized as follows:
\begin{itemize}
\item $\sigma_{\alpha}$ -  volatility-like parameters for market players’ vote in conditional probability to be bear in a bullish market.  
\item $k_{asym},\sigma_{\beta}=\sigma_{\alpha}*k_{asym}$ - volatility-like parameters for market players’ vote in conditional probability to be bear in a bearish market. 
\item $I_{low}$ - lower boundary of propagation in  $(\alpha,\beta)$ space  and  $\beta-\alpha > I_{low}$
\item $I_{up}$ - upper boundary of propagation in  $(\alpha,\beta)$ space  $\beta-\alpha < I_{up}$
\item $\mu$ - prime is known though it can be treated as a free parameter
\item $\Delta_{market}$ - lowering of implied volatility of the market players by market regulations. This parameter is not fitted but calculated at each run.
\item $\alpha_0$ - initial $\alpha$ coordinate of the population
\item $\beta_0$ -  initial $\beta$ coordinate of the population
\item $B$ - coefficient between votes and stock prices’ change
\end{itemize}

These parameters define the simulation and corresponding implied volatility surface $\sigma_{imp}(T,K)$. The implied volatility surface for specific parameters’ values is derived via a Monte Carlo simulation enforced by GPU processing\cite{Wang1995}. The resulting volatility surface is compared with the actual surface and multiple runs are made to adjust the parameters using a simplex optimization method\cite{Nelder1965} fro GNU Scientific Library\cite{Gough2009}.

\section{Results}

The results include sets of the model parameters together with the corresponding calculated volatility surfaces that provide the best match of the real implied volatility surfaces either of skew or smile types. Calibrating the model’s parameters against real market data accomplishes two main tasks. First, it demonstrates the model’s ability to describe a wide range of the real volatility surfaces for practical trading or hedging. Second, it identifies the main parameters of the model responsible for skew or smile volatility for better understanding of these phenomena.

Calibration is done either by using a single set of parameters' values for the entire implied volatility surface or by using different parameters' values for implied volatilities with different maturity times. The latter is done because we cannot fit all surface by a single set of parameters. Consequently, the parameters may be functions of maturity time, similar to the definition of implied volatility.

The model’s parameters were calibrated against real implied volatility surfaces corresponding to the S \& P index (SPX) at different times and Vodafone Company (VOD) stock: 
\begin{itemize}
\item SPX index at 12/26/01 with maturities of 1,3,6, and 12 months
\item VOD stock at 12/27/01 with maturities of 1,3,6, and 12 months
\item SPX index at 09/15/05 with maturities of 1 day together with 1,3,6,15 months
\end{itemize} 
These cases include implied volatilities with both smile and skew at different strengths.

The index is an average of many stocks; therefore, it represents well the average properties of the market. On the contrary, a single stock such as VOD can deviate from average market behavior and from general assumptions of the model.

The skew pattern of SPX was successfully fitted by a single set of parameters that are independent of time and price value. On the contrary, the smile pattern of VOD and early SPX cannot be fitted be a single set of parameters; therefore, it was fitted separately for each maturity time. 

\subsection{SPX 12/26/01}

\begin{figure}
  \centering
  \resizebox{0.5\textwidth}{!}{\includegraphics{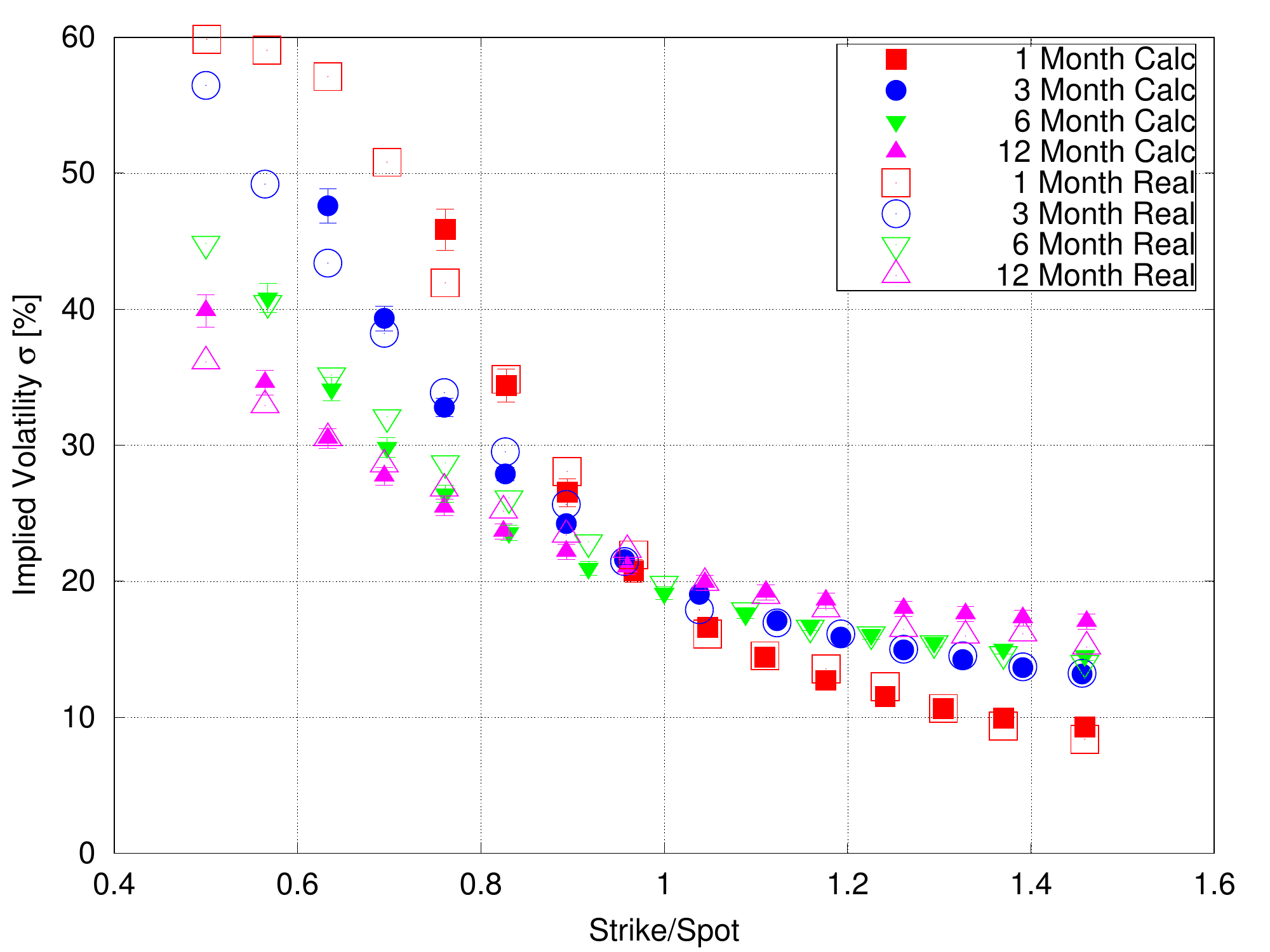}}
  \caption{{\bf Fit of the complete model to the implied volatility surface of SPX index from 12/26/01.} Calculated implied volatility surface following the optimized complete set of model parameters (see Table \ref{tab:spxcomp}) together with the real one. The error bars correspond to the accuracy of the parameters after multiple attempts of optimization.}
  \label{fig:fig10}
\end{figure}

The implied volatility surface of the SPX index from 12/26/01 demonstrates clear skew behavior at maturity times 1,3,6, and 12 months. First, we fit the surface with the entire set of model's parameters (see Fig. \ref{fig:fig10} and Table \ref{tab:spxcomp}). The results include both mean and standard deviation values obtained over many calibration runs. 

Small standard deviations (see  Table \ref{tab:spxcomp}) of the calibrated  herding limits, volatilities $\sigma_\alpha=\sigma$ and $\sigma_\beta=k_{asym}\sigma$, initial position of the community $(\alpha_0,\beta_0)$, and market volatility correction $\Delta_{market}$ indicate the important role of these parameters. The real value of prime $\mu=4.5\%$ at 12/26/01 is within the error range of the result. Interestingly, the initial position of the population is almost neutral $\alpha_0\approx\beta_o\approx 0.5$ and coefficient $B\approx 1$. 

\begin{table}[h]
\centering
\begin{tabular}{l|l}
 Parameter  &  Optimized value    \\
 \hline 
 $I_{up}$     &   0.24  $\pm$           0.009  \\
 $I_{low}$    &  -0.03  $\pm$          0.006  \\
 $\sigma$       &   1.37  $\pm$           0.04  \\
 $k_{asym}$      &   0.74  $\pm$           0.016  \\
 $\mu$      &   0.069  $\pm$           0.026  \\
 $\alpha_0$    &   0.50  $\pm$           0.004  \\
 $\beta_0$    &   0.51  $\pm$           0.005  \\
 $B$       &   1.09  $\pm$           0.03  \\
 $\Delta_{market}$     &   0.99  $\pm$           0.04  \\
\end{tabular}
\label{tab:spxcomp}
\caption{Parameters of the model (mean value $\pm$ standard deviation) that fit implied volatility surface of S \& P index from 12/26/01. All model’s parameters were calibrated, including the known prime that was $4.5\%$ at that time (12/26/01). Parameters were calibrated multiple times starting from different initial values. Error bars, therefore, also indicate the accuracy of the parameters. For instance, a high accuracy of the herding limits is essential to obtaining this result. The most surprising and interesting findings are: First, the allowed space of herding $I$ is shifted toward positive social influence. Second, the real value of prime 0.045 from 12/26/01 lies within the error range of the calculated value, though the model possesses no previous knowledge of its value. Third, $B$ is $\approx 1$. Fourth, the initial position of the population is $(\alpha\approx 0.5,\beta\approx 0.5)$. The latter, in the framework of the model, can be interpreted as neutral market condition where both increase and decrease of the prices is possible with equal probability.}
\end{table}

\begin{figure}
  \centering
  \resizebox{0.5\textwidth}{!}{\includegraphics{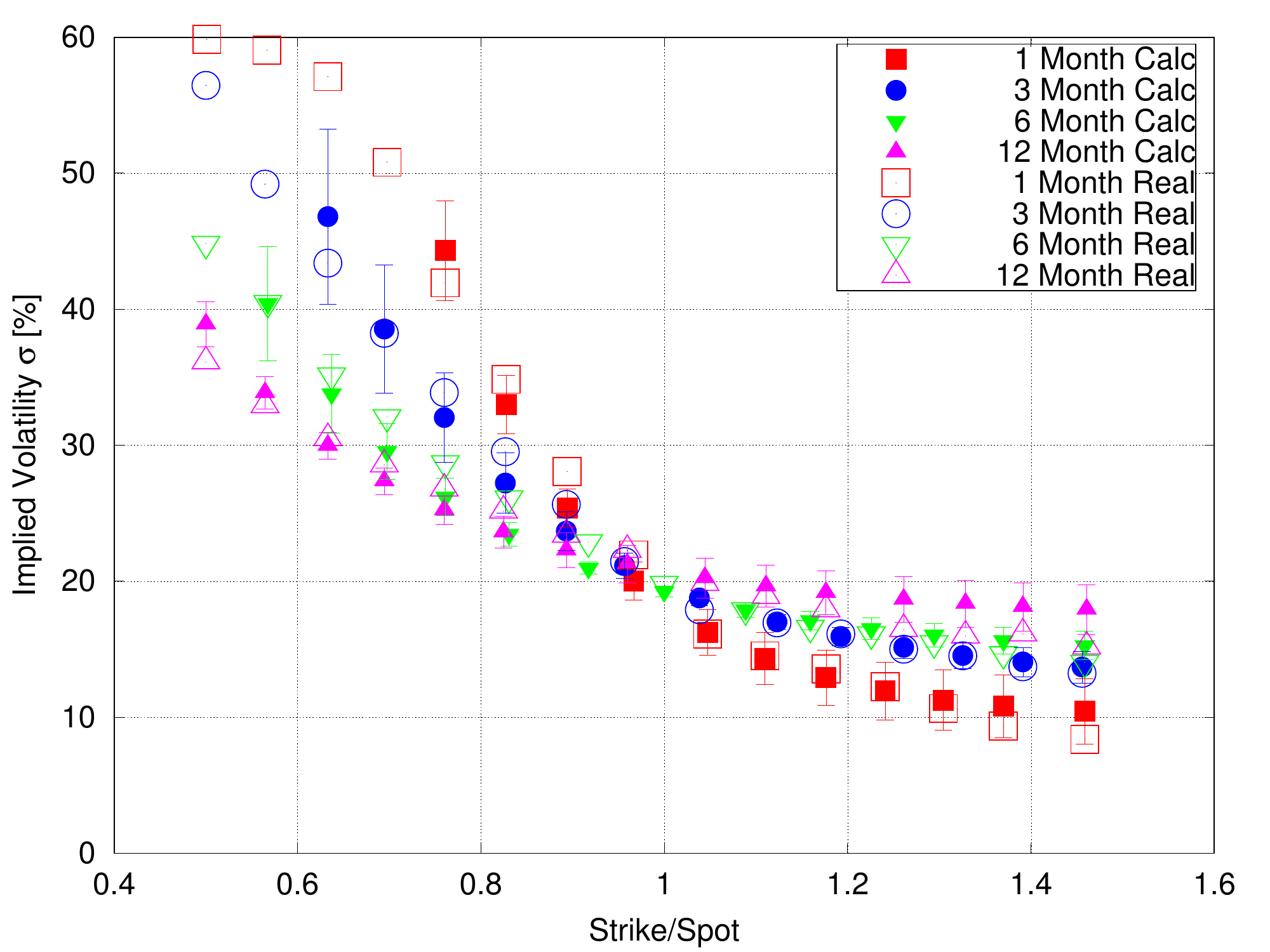}}
  \caption{{\bf The fit of the reduced model to the implied volatility surfaces of SPX index from 12/26/01.} The same as Figure \ref{fig:fig10}. The optimization was performed on the subset of the model parameters (see Table \ref{tab:5/tc}). The other parameters $B=1$, $(\alpha,\beta)=0.5$ following the complete optimization and supporting discussion. Prime was chosen as its value at that time. The accuracy is only slightly worse than in the case of the complete set of parameters.}
  \label{fig:fig10reduced}
\end{figure}

To highlight the most important parameters and to reduce the calibration time, some parameters where fixed with predefined values. Based on the previous results using the complete set of optimized parameters, the initial position is chosen to be $(\alpha_0=0.5,\beta_0=0.5)$ and $B=1$.  The value of prime $\mu=0.045$ corresponds to its real value at that time (12/26/01). This reduction barely affects the quality of the fit (see Fig. \ref{fig:fig10reduced} and Table \ref{tab:5/tc}).

\begin{table}[h]
\centering
\begin{tabular}{l|l}
 Parameter  &  Optimized value  \\
 \hline 
 $I_{up}$     &    0.23  $\pm$            0.03  \\
 $I_{low}$    &   -0.05  $\pm$            0.03  \\
 $\sigma$       &    1.38  $\pm$            0.13  \\
 $k_{asym}$      &    0.74  $\pm$            0.07  \\
 $\Delta_{market}$     &    0.88  $\pm$            0.05  \\
\end{tabular}
\caption{\label{tab:5/tc} Calibration of the model against implied volatility surface of S \& P index from 12/26/01 with a reduced set of parameters to simplify the model and reduce the calibration time. This set is chosen based on the results using the optimized complete set of parameters. The other parameters are assumed prime $0.045$, $B=1$, $(\alpha,\beta) = (0.5,0.5)$.}
\end{table}

\begin{figure}
\begin{center}
\resizebox{0.5\textwidth}{!}{\includegraphics{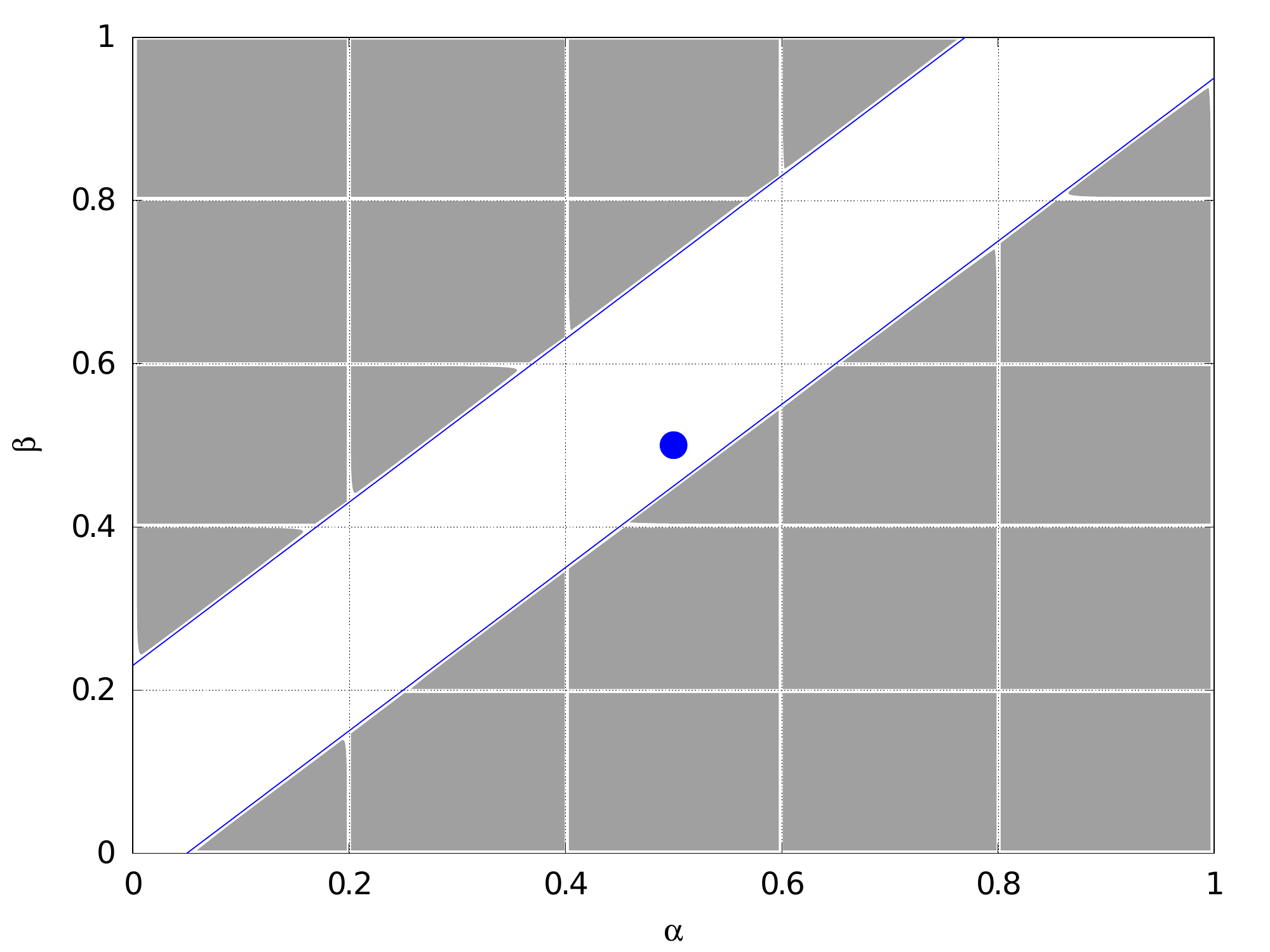}}
\caption{{\bf Herding limits obtained for SPX index from 12/25/01.} The dynamics of the market players starts at the point $(\alpha_0=0.5,\beta_0=0.5)$ and remains confined to the bright region; compare with Fig. \ref{fig7}. The limits impose boundaries on maximum positive (correlation with the market) and maximum negative (anti-correlation with the market) social influence.}
\label{fig8}
\end{center}
\end{figure}

The derived herding limits indicate that the decisions of the market players to be either bull or bear tend to correlate with market behavior. The allowed space for diffusion is shifted toward region $(\beta>\alpha)$ that includes state of the maximum correlation $(\alpha=0,\beta=1)$ (see Figs. \ref{fig8} and \ref{fig7}). On the other hand, there is no possibility of exact correlations with the market due to social influence limits. 

\subsection{VOD  12/27/01}

The implied volatility of the Vodafone Company (VOD) demonstrates clear smile behavior at maturity times of 1,3, and 6 months. The smile effect, however, reduces with greater maturity time. The implied volatility with maturity 1 year is of the skew type. As in the SPX case, the results are obtained over many calibration runs and include both the mean and the standard deviation value for optimized parameters. 

We found it impossible to fit the entire implied volatility surface of VOD with a single set of model parameters. Each maturity time then was fitted separately (see Table \ref{tab:vodnoilim}). For maturity times of 6 and 12 months, additional solutions were found (see Table \ref{tab:vodinfl}). Convergence depends on the initial choice of parameters.

\begin{figure}
  \centering
  \resizebox{0.5\textwidth}{!}{\includegraphics{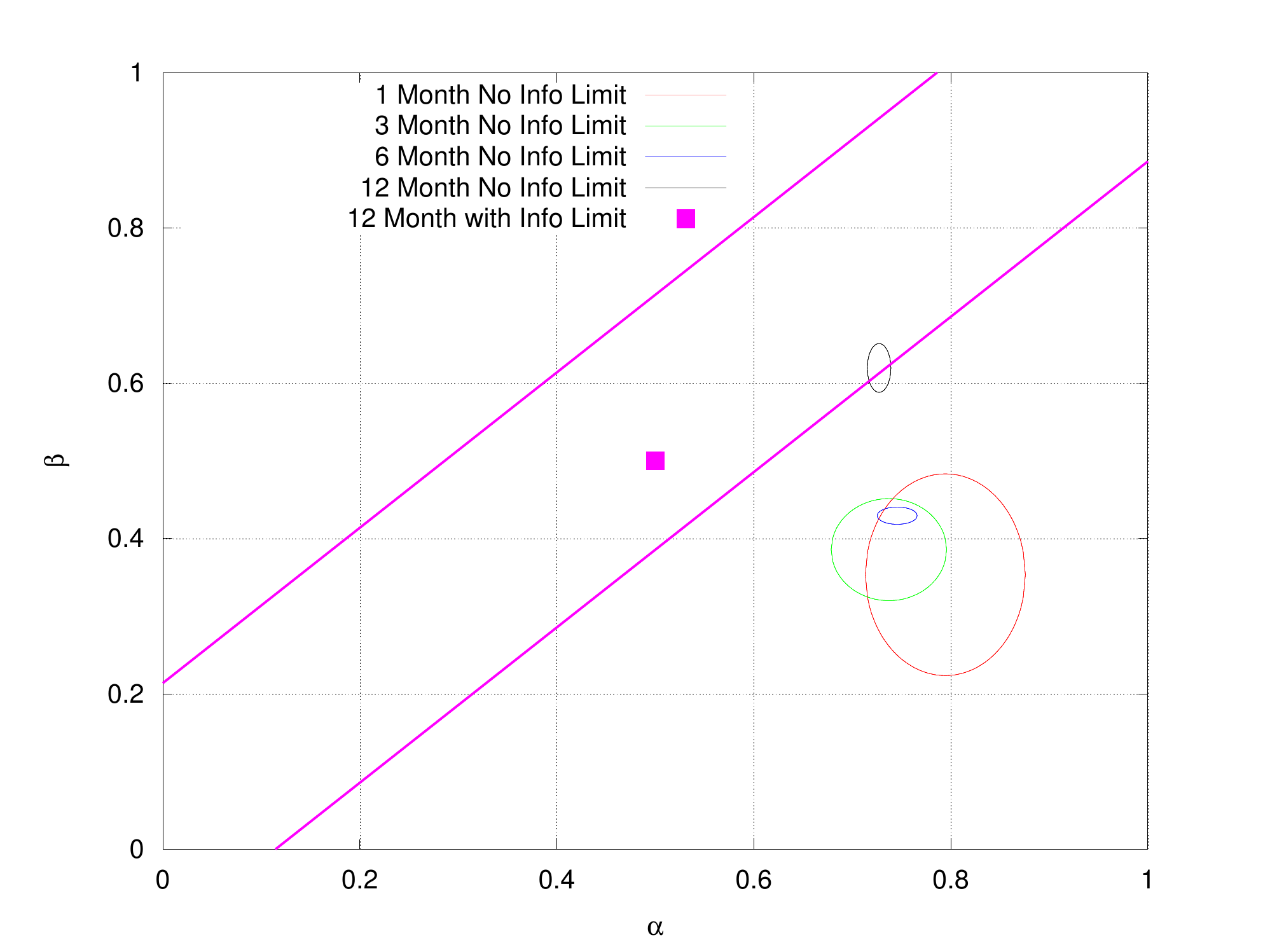}}
  \caption{{\bf The initial position of the population estimated separately for each maturity time of VOD volatility surface.} To demonstrate the change of the parameters with time the initial position with fit accuracy is shown for all four times from table \ref{tab:vodnoilim} together with the herding limits for 1 year maturity from \ref{tab:vodinfl}. One can argue that the population moves toward the center $(\alpha=\beta=0.5)$ and the affect of initial position is replaced by thr affect of herding limits. The later maturity times may be fitted either with the help of herding limits or, alternatively, with initial position that is different from neural community $(\alpha=\beta=0.5)$. The resolution of this question requires additional analysis.}
  \label{fig:fig11}
\end{figure}

The fit of strong smile pattern of single maturity time depends more on the initial position of the population, while moderate smiles and skew can be fitted both by initial condition and by herding limits (see Fig. \ref{fig:fig11}). The initial position may indicate a starting value of herding of the market players together with an excess of demand (disbalance between bears and bulls in the players’ community).

\begin{widetext}
\onecolumngrid
\begin{table}[h]
\begin{center}
\begin{tabular}{l|l|l|l|l}
 Parameter  &  1 Month                                &  3 Month                          &  6 Month                          &  1 Year                           \\
\hline 
$\sigma$       &  1.74 $\pm$          0.2  &  1.90 $\pm$          0.25  &  1.31 $\pm$           0.04  &  1.01 $\pm$           0.06  \\
 $k_{asym}$      &  0.64 $\pm$          0.25  &  0.83 $\pm$          0.084  &  0.85 $\pm$           0.03  &  0.91 $\pm$           0.01  \\
 $\alpha_0$    &  0.79 $\pm$          0.08  &  0.74 $\pm$          0.06  &  0.74 $\pm$           0.02  &  0.72 $\pm$           0.01  \\
 $\beta_0$    &  0.35 $\pm$          0.13  &  0.38 $\pm$          0.06  &  0.43 $\pm$           0.01  &  0.62 $\pm$           0.03  \\
 $B$       &  1.00 $\pm$          0.26  &  1.06 $\pm$          0.26  &  1.05 $\pm$           0.08  &  1.13 $\pm$           0.20  \\
 $\Delta_{market}$     &  0.17 $\pm$          0.22  &  0.44 $\pm$          0.1  &  0.21 $\pm$           0.06  &  0.29 $\pm$           0.10  \\
\end{tabular}
\end{center}
\caption{Parameters of the model (mean value $\pm$ standard deviation) that fit the implied volatility surface of VOD stock from 12/27/01. The fit was done without herding limits and with prime $=0.045$ at that time. No limits of herding were set due to the impossibility to fit the initial high smile with any herding limit. As in the case of SPX $B\approx 1$. The initial position, however, is drastically different from 0.5. The initial position, in the model’s framework, can be interpreted as a market in inertia either to increase or to decrease.}
\label{tab:vodnoilim}
\end{table}
\twocolumngrid
\end{widetext}

\begin{table}[h]
\begin{center}
\begin{tabular}{l|l|l}
 Parameter  &  6 Month                             &  1 Year                              \\
\hline
 $I_{up}$     &  0.29 $\pm$         0.06   &  0.21 $\pm$         0.03   \\
 $I_{low}$    &  -0.11 $\pm$         0.08  &  -0.11 $\pm$         0.02  \\
 $\sigma$       &  1.28 $\pm$         0.06   &  1.33 $\pm$         0.05   \\
 $k_{asym}$      &  0.91 $\pm$         0.04   &  1.07 $\pm$         0.03   \\
 $\Delta_{market}$     &  7.6e-03 $\pm$         5.1e-04   &  2.3e-03 $\pm$         5.8e-04   \\
\end{tabular}
\end{center}

\caption{Fit with herding limit when initial conditions for the population are around 0.5. The convergence is good. It is ambiguous whether the main phenomenon is initial position or herding limits.}
\label{tab:vodinfl}
\end{table}

There are multiple fit solutions for maturity times 6 and 12 months. A mild smile and an individual skew might be fitted either by herding limits or by initial position of the players’ community. The values of $\sigma$ are closer to each other in Table \ref{tab:vodnoilim} than in the initial position. This, together with SPX results and general reduction of social influence-like phenomena with time, favors the assumption of the herding limits solution for skew patterns.

The smile of implied volatility corresponds to a high value of the initial anti-herding between the market players, see \ref{fig:fig11}.

\subsection{SPX 09/15/05}

The SPX index 09/15/05 possesses a smile type of implied volatility for maturity times of 1 day and 1 month, together with a skew type for later maturity times\cite{Gatheral2006}. The results of parameter calibration are presented in Table \ref{tab:bookall}. The calibration of skew was done with initial conditions $(\alpha_0=0.5,\beta_0=0.5)$ to favor the influence of herding limits (see Fig. \ref{fig8}).

\begin{widetext}
\onecolumngrid
\begin{table}[h]
\begin{center}
\begin{tabular}{l|l|l|l|l|l}
 Parameter  &  1 Day                               &  1 Month                   &         3 Month                           &  6 Month                           &  15 Month                       \\
 $I_{up}$     &  1                                 &  1                               &  0.47 $\pm$           0.06   &  0.39 $\pm$           0.04   &  0.28 $\pm$           0.02   \\
 $I_{low}$    &            -1                      &  -1                              &  -0.20 $\pm$           0.05  &  -0.15 $\pm$           0.01  &  -0.11 $\pm$           0.01  \\
 $\sigma$       &  24.0 $\pm$         0.08   &  5.6 $\pm$          0.01  &  1.54 $\pm$           0.09   &  1.46 $\pm$           0.04   &  1.45 $\pm$           0.04   \\
 $k_{asym}$      &  0.97 $\pm$         0.039   &  0.62 $\pm$          0.02  &  0.61 $\pm$           0.08   &  0.80 $\pm$           0.04   &  0.87 $\pm$           0.03   \\
 $\alpha_0$    &  0.87 $\pm$         0.030   &  0.94 $\pm$          0.007  &  0.48 $\pm$           0.018   &  0.47 $\pm$           0.02   &  0.48 $\pm$           0.05   \\
 $\beta_0$    &  0.19 $\pm$         0.033   &  0.16 $\pm$          0.01  &  0.43 $\pm$           0.04   &  0.46 $\pm$           0.01   &  0.51 $\pm$           0.02   \\
 $B$       &  0.005 $\pm$         0.016   &  0.79 $\pm$          0.052  &  1.19 $\pm$           0.08   &  1.11 $\pm$           0.02   &  1.09 $\pm$           0.07   \\
 $\Delta_{market}$     &  -0.007 $\pm$         0.036  &  0.40 $\pm$          0.02  &  1.30 $\pm$           0.07   &  1.33 $\pm$           0.06   &  1.33 $\pm$           0.07   \\
\end{tabular}
\end{center}
\label{tab:bookall}
\caption{Model’s parameters(mean value $\pm$ standard deviation) that fit implied volatility surface of SPX index from 09/15/05. The fit was done without herding limits for early times and with herding limits for the later ones. The value of the prime was $=0.065$ at that time. Parameter $B\approx 1$ for later times, though, is different from the initial ones. The initial position $(\alpha_0,\beta_0)$ is drastically different from 0.5. This case can be considered as a mix of the previous ones.}
\end{table}
\twocolumngrid
\end{widetext}

\begin{figure}
  \centering
  \resizebox{0.5\textwidth}{!}{\includegraphics{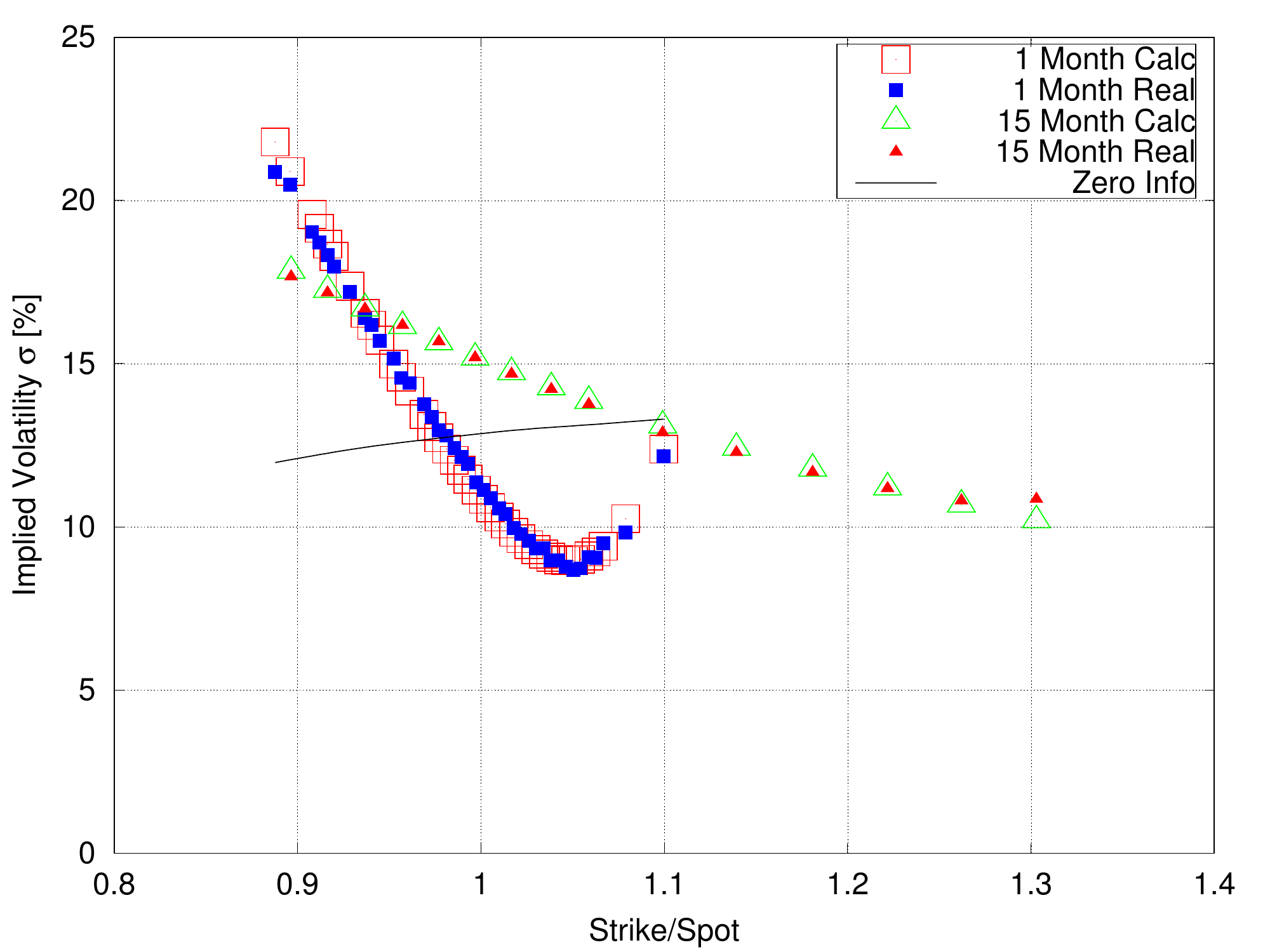}}
  \caption{{\bf Fit of smile and skew implied volatilities.} The fit of 1 month and 15 month implied volatilities of SPX index 09/15/05. For comparison, we added a calculated implied volatility at Black-Scholes limit with $I_{up},I_{low}= \pm 0.01$. This graph demonstrates three types of implied volatilities.}
  \label{fig:fig17}
\end{figure}

The fit of implied volatilities with 1 month and 15 month maturity times is presented in Fig. \ref{fig:fig17}. The fit of the smile mainly depends on the initial position of the market players. The skew depends more on imposed herding limits. A stronger limit can bring implied volatility to almost flat Black-Scholes form.

\begin{figure}
  \centering
  \resizebox{0.5\textwidth}{!}{\includegraphics{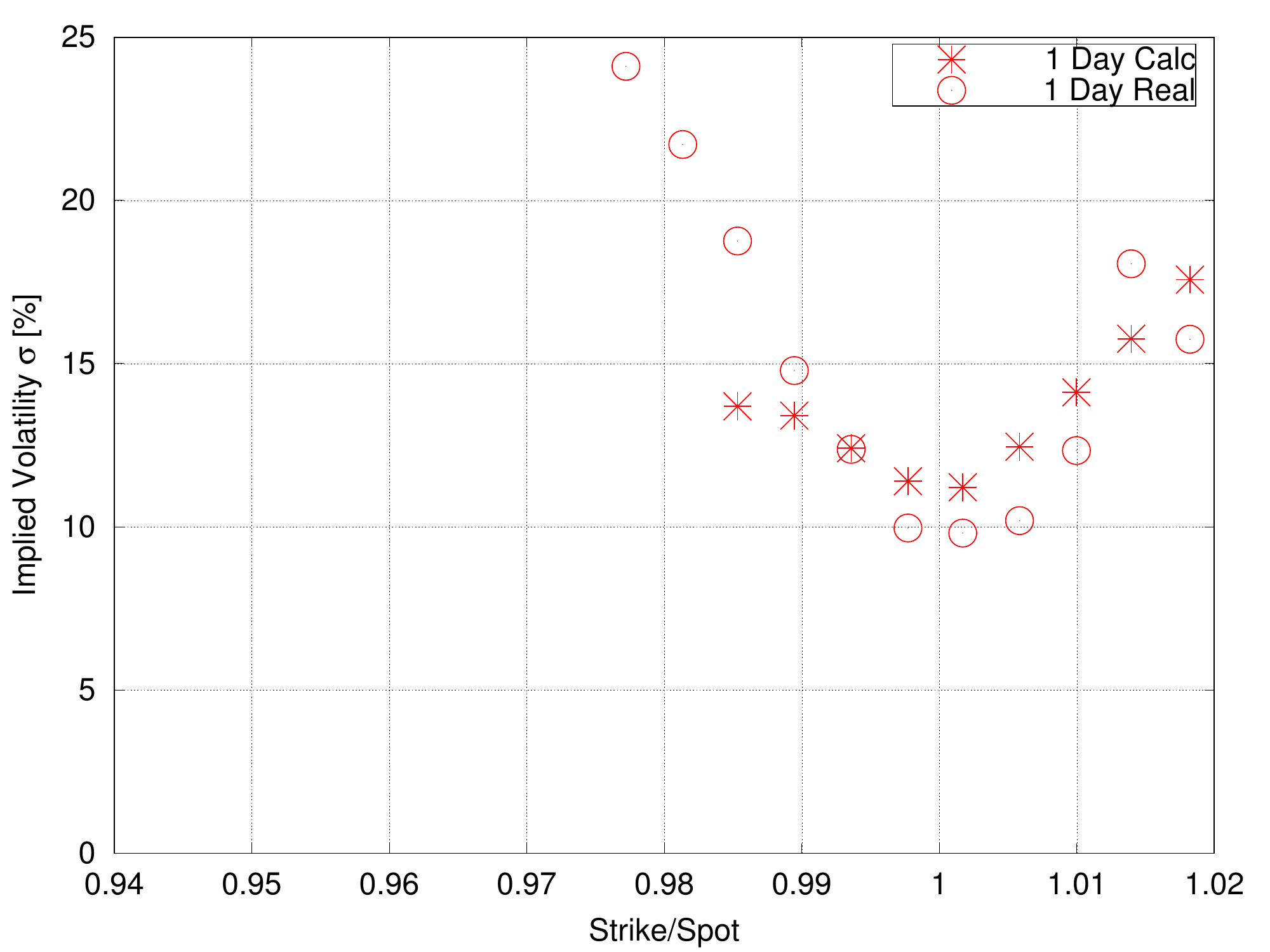}}
  \caption{{\bf The fit of implied volatility surface with 1 day maturity time SPX index 09/15/05} Implied volatility is characterized by strong smile. The fit is valid only for $Strike/Spot \approx 1$ since for short maturity times, the probabilities of the price distribution are close to initial $\delta$ function at $Strike/Spot=1$.}
  \label{fig:fig12}
\end{figure}

The strong smile at 1 day maturity can be fitted (see Fig. \ref{fig:fig12}), though it requires extreme values of volatility relative to other maturity times (see Table \ref{tab:bookall}. The initial position $(\alpha_0,\beta_0)$ indicates that the market players anti-correlated with the market.  

\begin{figure}
  \centering
  \resizebox{0.5\textwidth}{!}{\includegraphics{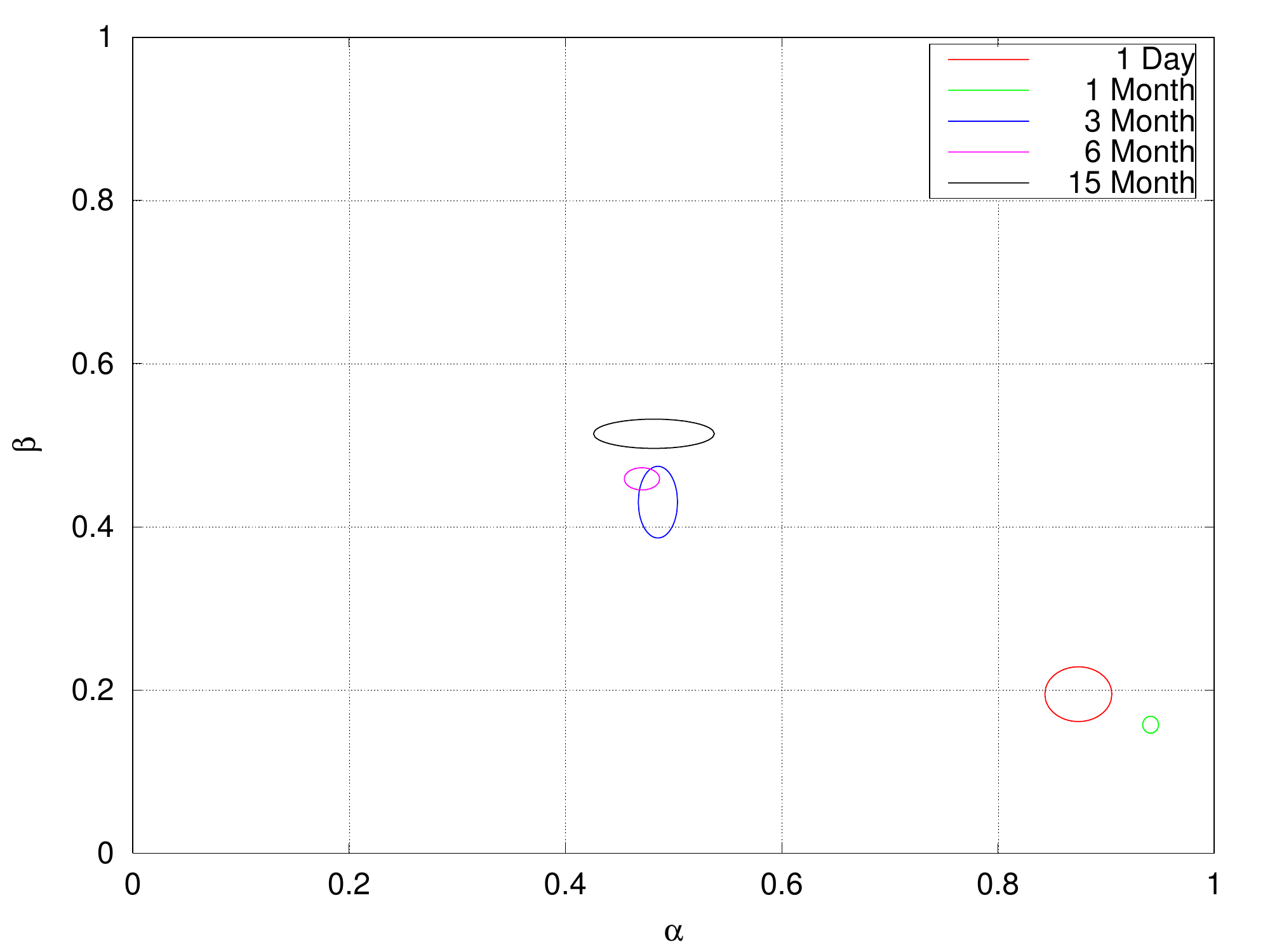}}
  \caption{{\bf The initial position of population as a function of maturity time for SPX index of 09/15/05}. The calibrated initial positions (see Table \ref{tab:bookall}) of market players’ community $(\alpha_0,\beta_0)$ are presented as ellipses in  $(\alpha,\beta)$ space. The center of the ellipse corresponds to the mean value while its axes are standard deviations of calibration over numerous runs. Early maturity times require a high value of initial social influence or deviation from the Black-Scholes limit $\alpha\approx \beta$, see Figure \ref{fig5}. The results for later time assume almost no initial social influence in the community.}
  \label{fig:fig14}
\end{figure}

Following results \ref{tab:bookall}, the initial position of the community converges to the region of small social influence with maturity time (Fig. \ref{fig:fig14}). It is clearly separated into two groups of early and later maturity times.   

\begin{figure}
  \centering
  \resizebox{0.5\textwidth}{!}{\includegraphics{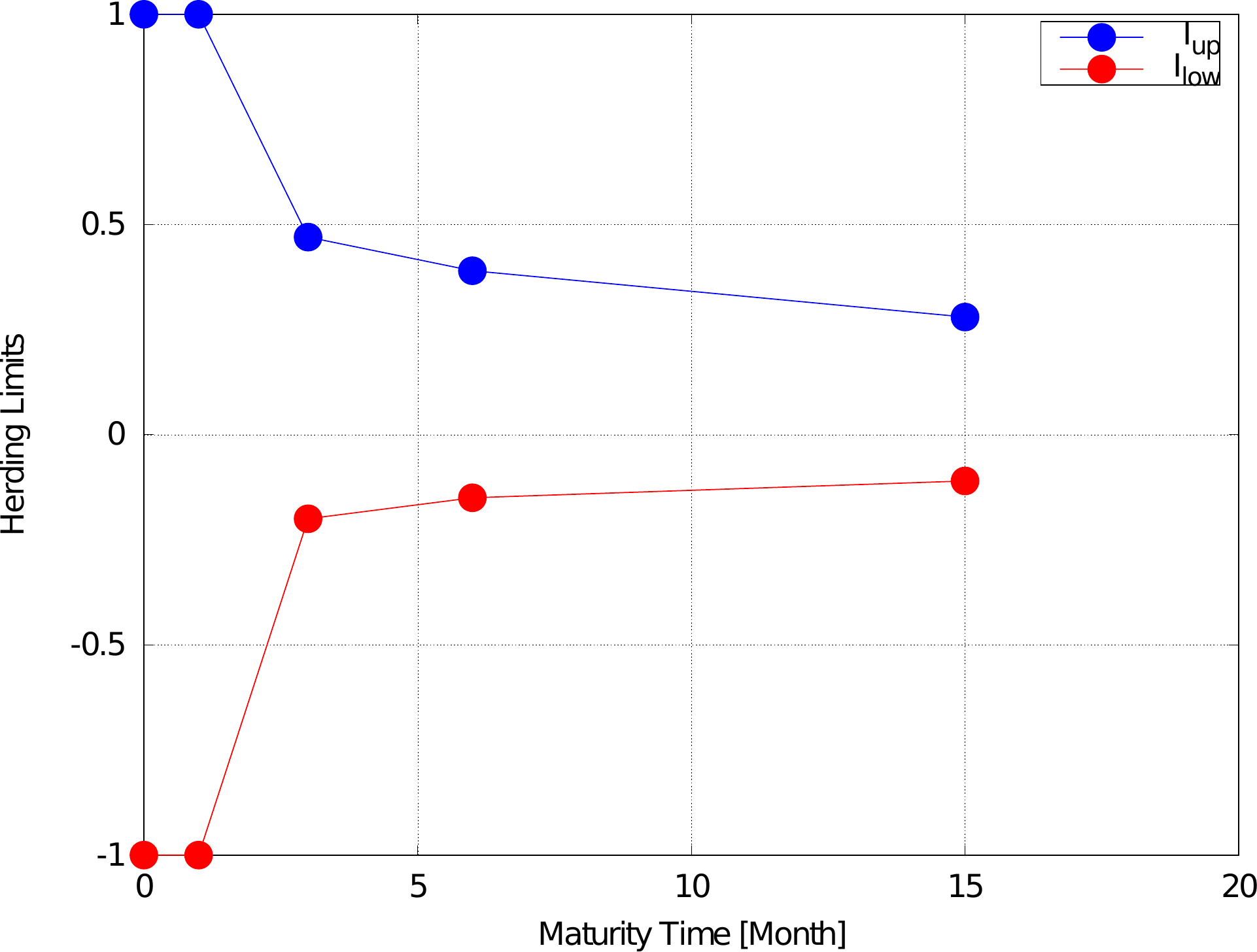}}
  \caption{{\bf The limits $I_{up}$ and $I_{low}$ as the functions of maturity time for SPX index from 09/15/05.} The $I_{up}$ and $I_{low}$ limit possible correlation and anti-correlation of the players with the market. The limits were calibrated separately for each maturity time (see Table\ref{tab:bookall}). Implied volatilities for short maturity times of 1 day and 1 month possess strong smile form. The fit of these volatilities requires lack of the herding limits $(I_{up}>1,I_{low}<-1)$. For later times herding boundaries gradually reduce to the direction of the Black-Scholes limit. The allowed region between two limits is shifted toward a positive correlation between market players’ opinions and market players’ dynamics.}
  \label{fig:bookI}
\end{figure}

Herding limits as a function of time according to Table \ref{tab:bookall} are shown in Fig. \ref{fig:bookI}. The limits converge to $0$ with time and bring the market players’ community closer to the Black-Scholes limit. The lower absolute value of $I_{low}$ than $I_{up}$ at later maturity times indicates a positive correlation between the market behavior and players’ predictions.

Market players’ dynamics, following the results for SPX indices and VOD stock, at early maturity times, may possess a high value of social influence. This is reasonable, since information about the current state of the community is preserved for a short time into the future. Longer maturity times are characterized by strong herding limits. The initial position of the population lacks almost any social influence value. Indeed, no memory is preserved for a long time and, consequently, the market envisioned by its players converges to the Black-Scholes limit with maturity time.

\section{Discussion}

To corroborate the basic assumptions of the model, let us compare the implied volatility surfaces of the SPX index, the VOD stock, and the results of this work. The model, after appropriate calibration of its parameters, successfully fits real implied volatility surfaces of both smile and skew types. The model’s parameters describe the community of the market players together with the market that transforms the players’ action into observable prices. The obtained description of the market, together with players’ community, is analyzed in light of the model assumptions and common sense.

The assumption of the Black-Scholes (log normal) price dynamics in the case of random (zero social influence) acts of the market players is supported both by properties of real implied volatility surfaces and the results of this work. The properties of market players for the purpose of evaluation options with greater maturity times are less correlated with their current values. Indeed, implied volatilities converge to the Black-Scholes “flat” form with maturity time. The same happens for calibrated herding limits that converge to their Black-Scholes limit $I_{up}\approx I_{low}\approx 0$ with maturity time (see Figure \ref{fig:bookI}). 

The assumptions of market mechanism in the form of deterministic function (without a stochastic term) are supported by the ability to fit real implied volatility surfaces. Each parameter of the model is justified and essential. Moreover, the model requires an additional heuristic parameter $\Delta_{market}$ to match reality. This parameter, however, is justified as market's reduction of market players’ volatility by regulations for stability and crisis prevention. 

The distinctive explanations for skew and smile phenomena as limits and initial values of herding, respectively, agree with previous association of skew and smile implied volatility surfaces with indices and stocks. The index is an average price of multiple stocks and, therefore, better presents the average properties of the market. Consequently, in the case of indices, the limits on herding should be stronger and the dependence on initial value of herding is low. Contrary, individual stock might possess broader limits on herding and its value for short maturity times. 

Upper $I_{up}$ and lower $I_{low}$ limits on herding correspond to the maximum of the correlation and anti-correlation of market players’ opinions, respectively. The greater $I_{up}$ the greater the possibility of collective phenomena of the players.

The definition of market player strategies using conditional probabilities stems from similar techniques of the game theory\cite{Nowak2004}. The question of optimal strategy and its evolutionary stability remains under discussion\cite{Press2012}. An analysis of the market may contribute to this discussion due to the huge amount of available data. On the other hand, there is also an interesting discussion whether evolution is relevant for the market players\cite{Farmer2002a}\cite{Cruz2013}\cite{Gao2013}\cite{Shi2013}\cite{Durrett2012}\cite{Zhu2011}\cite{Dong2007}.

The model’s limitations and drawbacks are the following: the transaction costs and other players’ states but bull or bear were not taken into account; equal financial weight of the players was assumed; and the prime interest rate was omitted in the voting process of the market players and considered only as monetary value change during calculating implied volatility. The prime converges close to its real value, however, if assumed to be a free parameter. In addition, we did not take into account near neighbors’ topological constraint. A single market player may affect any other player with a limit on the total amount of social influence it generates. The excessive communication abilities of the present time and a good match of the model results with real prices justify this approach.

This work predicts the stochastic process for price dynamics that depends on a small number of well-justified parameters. It is an advantage over approaches of local\cite{Derman1994} and stochastic volatilities\cite{Hull1987}. The latter together with completely heuristic descriptions of implied volatility may be advantageous for practical needs by their speed. Our model, however, can be accelerated by further massive parallelization and, in general, is comparable to any Monte-Carlo based trading tools.

The results of this work are potentially relevant for any agent-based simulation\cite{Feng2012}\cite{Bertella2014}\cite{Kim2014}\cite{Liu2014}\cite{Chen2014}\cite{Wei2013}\cite{Schmitt2012}\cite{Chakraborti2011} because the model’s major assumptions are independent of the market microstructure. Moreover, the model is impervious to modifications. The need for a function that transforms the voting process to log normal requires the terms $\mu$ and $\sigma$ to depend on each other and prevents their arbitrary modifications. Moreover, neglecting (\ref{gam1}) and writing the process in the form $d\gamma\propto (1-2\gamma)dt+2\sqrt{\gamma(1-\gamma)}dz$ leads to unrealistic bounded expression for the price–vote relation $F\propto \arcsin (2\gamma-1)$ instead of (\ref{F0}).

The results of this work support the relation of information technologies (IT) contribution to the financial crisis of 1987. The implied volatilities of the SPX index acquired significant skew during this crisis\cite{Rubinstein1994}. According to our model, it indicates the growth of possible social influence of the market players. The latter might be a consequence of extensive IT technology modification of the markets and trading at that time. It corroborates that information-like phenomena are the cause rather than the consequence of the crisis.

The herding of the market players might be a new tradable parameter, similar to the volatility index (VIX)\cite{Brenner1989}, which describes the past values of market's volatility. On the contrary, in this work, herding describes possible future developments of the market players’ community. It may be an important market indicator for crisis analysis\cite{Sornette2003}. Moreover, herding can be estimated by other means from internet responses and artificial market experiments for comparison with the market, or even for predicting the market behavior.

\section{Conclusion}

We presented a framework for modeling the market prices’ dynamics based on opinion dynamics and herding in the trading community. This framework uses the social influence and mutual information between the players as a quantitative measure of the herding effect and corresponding deviation of the prices from the Black-Scholes model. The derived relation between opinion dynamics and price formation is general and argued to be independent of exact market mechanism. The calculated option prices fit real market data and can be useful for trading and hedging. In addition, the estimated herding from market data can be compared with herding from other sources such as artificial markets, news, or social networks.


\end{document}